\documentstyle[aps,prl,preprint,floats,epsfig]{revtex}  
\begin{document}
%\doubles

\preprint{\vbox{ \hbox{ BELLE-CONF-0114}}}
%\twocolumn[\hsize\textwidth\columnwidth\hsize\csname 
%@twocolumnfalse\endcsname 

\title{
 \quad\\[1cm] \Large
Study of three-body charmless $B$ decays at Belle}

\author{The Belle Collaboration}

\maketitle

% to make it single spaced
\tighten

\begin{abstract}

    Using a data sample of $21.3$ fb$^{-1}$ collected by the Belle detector,
three body charmless decays $B^+\to K^+h^+h^-$ have been studied. With no
assumptions on the intermediate mechanisms, the following branching
fractions have been measured for the first time
$ {\cal{B}}(B^+\to K^+\pi^-\pi^+)~~= (58.5\pm7.1\pm 8.8)\times10^{-6}$ and 
$ {\cal{B}}(B^+\to K^+K^-K^+)     = (37.0\pm3.9\pm 4.4)\times10^{-6}$.
We also present the first observations of the decay mode
$B^+\to K^{*0}(892)\pi^+$ with a branching fraction of
${\cal{B}}(B^+\to K^{*0}(892)\pi^+) =
(16.7^{+3.7+2.1+3.0}_{-3.4-2.1-5.9})\times10^{-6}$
and the decay mode $B^+\to f_0(980)K^+$ with a product branching fraction of
${\cal{B}}(B^+\to f_0(980)K^+)\times{\cal{B}}(f_0(980)\to \pi^+\pi^-) =
(11.7^{+2.5+1.5+4.1}_{-2.7-1.5-1.0})\times10^{-6}$.

\end{abstract}
\pacs{PACS numbers: 13.20.H }  

{\renewcommand{\thefootnote}{\fnsymbol{footnote}}

%% >>>>>> LP2001 authorlist will go here
%% >>>>>> current version is /g8home/browder/lp2001/authorlist_lp2001.tex

\begin{center}
  K.~Abe$^{9}$,               % KEK
  K.~Abe$^{37}$,              % TohokuGakuin
  R.~Abe$^{27}$,              % Niigata
  I.~Adachi$^{9}$,            % KEK
  Byoung~Sup~Ahn$^{16}$,      % Korea
  H.~Aihara$^{39}$,           % Tokyo
  M.~Akatsu$^{20}$,           % Nagoya
  K.~Asai$^{21}$,             % Nara
  M.~Asai$^{10}$,             % Hiroshima
  Y.~Asano$^{44}$
% ...
\end{center}

\small
\begin{center}
$^{1}${Aomori University, Aomori}\\
% ...
$^{9}${High Energy Accelerator Research Organization (KEK), Tsukuba}\\
$^{10}${Hiroshima Institute of Technology, Hiroshima}\\
% ...
$^{13}${Institute of High Energy Physics, Vienna}\\
% ...
$^{16}${Korea University, Seoul}\\
% ...
$^{20}${Nagoya University, Nagoya}\\
$^{21}${Nara Women's University, Nara}\\
% ...
$^{27}${Niigata University, Niigata}\\
% ...
$^{37}${Tohoku Gakuin University, Tagajo}\\
% ...
$^{39}${University of Tokyo, Tokyo}\\
% ...
$^{44}${University of Tsukuba, Tsukuba}\\
% ...
\end{center}

\normalsize

\setcounter{footnote}{0}
\newpage

\normalsize

%\onecolumn

\vspace*{0.5cm}

\section{Introduction}

   During the last few  years, a considerable amount of new information
on charmless hadronic decays of B-mesons has been reported, primarily
by the CLEO Collaboration.  The discoveries of the $B \to K\pi$ 
and $B \to \pi\pi$ decay modes~\cite{cleokpi} have provided 
a real basis for searches for direct-CP violating 
effects in the B-meson system. 

However, because of large combinatoric backgrounds,  
studies of charmless B decays have concentrated mainly
on two-body decay processes.  
In this paper, we report the results of a study 
of decays $B^+ \to K^+h^+h^-$ 
($h$ stands for a charged pion or kaon) where no  assumptions are
made about intermediate hadronic resonances. 
The inclusion of charge conjugate states is implicit 
throughout this report unless explicitly stated otherwise.

The data sample used for this analysis 
consists of 21.3 fb$^{-1}$ taken at the $\Upsilon(4S)$
(on-resonance) and 2.3 fb$^{-1}$ taken 60 MeV below for continuum studies  
(off-resonance).  The data were collected with the Belle detector~\cite{NIM} 
operating at the KEKB asymmetric energy $e^+e^-$ collider~\cite{KEKB}.

%%%%%%%%%%%%%%%%%%%%%%%%%%%%%%%%%%%%%%%%%%%%%%%%%%%%%%%%%%%%%%%%%%%%%%%%%%%%%

\section{The Belle detector}

Belle is large-solid-angle spectrometer based on a 1.5~Tesla superconducting
solenoid magnet.  Charged particle tracking is provided by a
silicon vertex detector (SVD) and a cylindrical drift chamber
(CDC) that surround the interaction region. The SVD consists
of three approximately cylindrical layers of double-sided silicon 
strip detectors; one side of each detector measures the $z$ coordinate 
and the other the $r$-$\phi$ coordinate.
The CDC has  50 cylindrical layers of anode
wires; the inner three layers have instrumented cathodes
for $z$ coordinate measurements~\cite{CDC}.
Twenty of the wire layers are inclined at small angles 
to provide small-angle stereo
measurements of $z$ coordinates along the particle trajectories.
The charged particle acceptance covers the laboratory polar angle between
$\theta=17^{\circ}$ and $150^{\circ}$ corresponding to about 92\%
of the full solid angle in the CMS.

Tracks are fit using an incremental Kalman filtering technique,
where individual measurements found by the CDC pattern recognition
algorithm  are added successively to update the track's parameters
and  covariance matrix at each measurement surface.
This approach to track fitting minimizes the effects of multiple 
Coulomb scattering on the determination of the track parameters.
Hits from the SVD are associated and included during the last 
steps of this recursion. The momentum resolution  is determined from 
cosmic rays and $e^+ e^-\to\mu^+\mu^-$ events to be 
$\sigma_{p_t}/p_t = (0.30 \oplus 0.19 p_t)\%$, where $p_t$ is the 
transverse momentum in GeV/$c$.

  Charged hadron identification is provided by $dE/dx$ measurements 
in the CDC,  a mosaic of 1188 aerogel \v{C}erenkov counters (ACC), 
and a barrel-like array of 128 time-of-flight scintillation counters 
(TOF).  The $dE/dx$  measurements have a resolution for hadron tracks 
of 6.9\% and are useful for $\pi /K$ separation for $p_{lab} < 0.8$~GeV/$c$
and $p_{lab} > 2.5$~GeV/$c$ where $p_{lab}$ is the  laboratory momentum.
The TOF system has a time resolution for hadrons that is 
$\sigma \simeq 100$~ps and provides $\pi /K$ separation for  
$p_{lab} < 1.5$~GeV/$c$~\cite{TOF}. The indices of refraction of the ACC 
elements vary with polar angle to match the kinematics of the
asymmetric energy environment of Belle and cover the  range 
$1.5 < p_{lab} < 3.5$~GeV/$c$~\cite{ACC}. 

  Hadron identification is accomplished by combining
the information from these three subsystems into a single 
number using the likelihood method:
\[ {\cal{L}}(h) = {\cal{L}}^{ACC}(h)\times {\cal{L}}^{TOF}(h)\times{\cal{L}}^{CDC}(h),\]
where $h$ stands for the hadron type ($\pi$, $K$, $p$).
High momentum tagged kaons and pions from
kinematically selected $D^{*+}\to D^0\pi^+$; $D^0\to K^-\pi^+$
decays are used to determine a charged particle identification
efficiency of about 90\% and a misidentification probability of about~8\%.

   Electromagnetic showering particles are detected in an array of 8736 
CsI(Tl) crystals located in the magnetic volume and covering the same solid 
angle as the charged particle tracking system~\cite{CSI}.  The energy 
resolution for electromagnetic showers is 
$\sigma_E/E = (1.3 \oplus 0.07/E \oplus 0.8/E^{1/4})\%$, ($E$ in GeV). 
Neutral pions are detected via their $\pi^0\to\gamma\gamma$ decay. 
The $\pi^0$ mass resolution varies slowly with energy, averaging 
$\sigma_{m_{\pi^0}} = 4.9$~MeV.

   Electron identification in Belle is based on a combination of $dE/dx$ 
measurements in the CDC, the response of the ACC, and the position, shape
and total energy (i.e. $E/p$) of the shower registered in the CsI 
calorimeter. 
The electron identification efficiency, determined by embedding Monte
Carlo tracks in multihadron data, is greater than 92\% for tracks with
$p_{lab}>1.0$~GeV/$c$ and the hadron misidentification probability,
determined from $K_S\to \pi^+\pi^-$ decays, is below 0.3\%.

  The 1.5~T magnetic field is returned via an iron yoke that is instrumented 
to detect muons and $K_L$ mesons.  This detection system, called the KLM,
consists of alternating layers of charged particle detectors and
4.7~cm thick steel plates. The total steel thickness 
of 65.8 cm plus the material of the inner detector corresponds to 4.7 nuclear
interaction lengths at normal incidence. The system covers 
polar angles between $\theta=20^\circ$ and $155^\circ$ and 
the overall muon identification efficiency, determined by a track
embedding study similar to that used in the electron case, is greater
than 87\% for tracks reconstructed in the CDC with $p_{lab} > 1.0$~GeV/$c$.
The corresponding pion misidentification probability
determined from $K_S\to \pi^+\pi^-$ decays is less than 2\%.

%%%%%%%%%%%%%%%%%%%%%%%%%%%%%%%%%%%%%%%%%%%%%%%%%%%%%%%%%%%%%%%%%%%%%%

\section{Event selection}

  Charged tracks are required to satisfy a set of track quality cuts
based on the average hit residual and the impact parameters in both the 
$r$-$\phi$ and $r$-$z$ planes. We require that the transverse track momenta be
greater than 100 MeV/$c$ to reduce low momentum combinatoric background.
All the cuts used for the selection of charged tracks are listed in 
Table~\ref{trkcuts}.

\begin{table}[htb]
\caption{Parameters used for selection of charged tracks.}
\medskip
\label{trkcuts}
  \begin{tabular}{ll} 
   Parameter  \hspace*{9.8cm}          &  Cut value                         \\ \hline 
   Transverse momentum, $p_t$          &  $p_t > 0.1$~GeV/c                 \\ 
   Impact parameter, $\Delta R$        &  $|\Delta R| < 0.25$~cm            \\ 
   Impact parameter, $\Delta Z$        &  $|\Delta Z| < 2.50$~cm            \\ 
   Number of CDC axial  hits, $N_{ah}$ &  $N_{ah} > 10$                     \\ 
   Number of CDC stereo hits, $N_{sh}$ &  $N_{sh} > 5$                      \\ 
  \end{tabular}
\end{table}

Charged particles are 
identified as  $K$'s or $\pi$'s by cutting on the likelihood ratio (PID): 
\[ PID(K) = \frac{{\cal{L}}(K)}{{\cal{L}}(K)+{\cal{L}}(\pi)};
PID(\pi) = \frac{{\cal{L}}(\pi)}{{\cal{L}}(\pi)+{\cal{L}}(K)} = 1 - PID(K)\]
  At large momenta ($>$2.5 GeV/$c$) only the ACC and $dE/dx$ are used 
since here the TOF provides no significant separation of kaons and pions.
For all three-body final states except the $KKK$ final state the likelihood
ratio  for kaon candidates is required to be greater than 0.8.
Otherwise the charged track is identified as a pion. 
For the $KKK$ final states, we require $PID(K)>0.5$ to maintain
high efficiency.

  All charged tracks are also required to satisfy an
electron veto requirement, which demands
that the electron likelihood is less than 0.95.
In addition, all charged kaon candidates are required to satisfy a
proton veto:

\[ PID(p) = \frac{{\cal{L}}(p)}{{\cal{L}}(p)+{\cal{L}}(K)} <0.95 \]

   The candidate events are identified by using the beam-constrained mass 
$M_{BC}= \sqrt{s/4 - P_B^{*2}}$ 
and the  energy difference $\Delta E = E_B^* - \sqrt{s}/2$,
where $E_B^*$ and $P_B^*$ are the measured energy and 3-momentum of the $B$ 
candidate in the $\Upsilon (4S)$ rest frame and $\sqrt{s}$ is the total 
energy in $\Upsilon (4S)$ rest frame.

   The $M_{BC}$ signal distribution  is well modeled by 
a single Gaussian function. The central value and width of the 
Gaussian function are determined from the 
$B^+\to \bar{D^0}\pi^+$, $\bar{D^0}\to K^+\pi^-$ signal to be
5.2805 GeV/$c^2$ and 2.75 MeV/$c^2$ respectively.
The $M_{BC}$ resolution is primarily due to the energy spread of the
$e^+$ and $e^-$ beams and is found to be
independent of the particular three charged track final state.
We use the ARGUS function~\cite{ARGUS} to describe the background in the
$M_{BC}$ distribution: $f(M_B) = \sqrt{1 - x^2}\exp[-\xi(1-x^2)]$,
where $x = M_B/E_{beam}^*$ and $\xi$ is a free fit parameter.

   The $\Delta E$ signal shape is parameterized 
by the sum of two Gaussians with
the same mean. The shape of the background in the 
$\Delta E$ distribution from non-resonant $e^+e^-\to q \bar{q}~(q=u,d,s,c)$
continuum events is parameterized
by a first order polynomial.
The $\Delta E$ shape due to background from other 
$B$ and $\bar{B}$ decay processes  has a substantial dependence 
on the final state being studied. 
Unless explicitly stated otherwise, the contributions 
from these  backgrounds are also parameterized by a first order polynomial.

  In the following, we refer to the ``$B$ signal region;'' this is
defined as: 
\begin{center}
$5.272<M_{BC}<5.289$ GeV/$c^2$;~$|\Delta E|<40$~MeV.
\end{center}

%%%%%%%%%%%%%%%%%%%%%%%%%%%%%%%%%%%%%%%%%%%%%%%%%%%%%%%%%%%%%%%%%%%%%%

\section{Background suppression}

   An important issue for this analysis is the suppression of the large
combinatoric background that is dominated by $q\bar{q}$ continuum
events.  To suppress this background, we use a set of variables 
that characterize the event topology.

   Since the two $B$ mesons produced from the $\Upsilon (4S)$ decay are 
nearly at rest in the CMS frame, the angles of the decay products
of two $B$'s are uncorrelated and the events tend to be spherical. 
In contrast,
hadrons from continuum $q\bar{q}$ events tend to exhibit a two-jet 
structure.
Figure~\ref{shape}a shows distributions of $|\cos(\theta_{Thr})|$, 
where $\theta_{Thr}$ is the angle between the thrust axis of 
the $B$ candidate and that of the rest of the event.
The distribution is strongly
peaked near
$|\cos(\theta_{Thr})|\simeq 1.0$ for $q\bar{q}$ 
events while it is nearly flat for $B\bar{B}$ events.
We require $|\cos(\theta_{Thr})|<0.80$ for all modes under consideration;
this eliminates 83\% of the continuum background and 
retains 79\% of the signal events. 

   After the imposition of the 
$\cos(\theta_{Thr})$, $q\bar{q}$ and $B\bar{B}$
requirements, the remaining events 
still have some differences in topology that 
are exploited for further continuum suppression. 
We construct a ``Virtual Calorimeter''~\cite{VCal} 
by dividing the space around the candidate thrust axis into nine polar angle 
intervals of $10^\circ$ each; the $i$-th
interval covers angles from
\mbox{($i$ - 1)$\times 10^\circ$} to \mbox{$i \times 10^\circ$}.
We define the momentum flows, $x_i (i = 1,9)$, into the $i$-th interval as a 
scalar sum of the momenta of all charged tracks and neutral showers directed 
in that interval. The momentum flow in corresponding 
forward and backward intervals
are combined.

   Angular momentum conservation  provides some additional 
discrimination between
$B\bar{B}$ and continuum $q\bar{q}$ events. In $q\bar{q}$ production, 
the direction of the 
candidate thrust axis, $\theta_{T}$, with respect to the beam axis in 
the cms frame tends to maintain the 
$1 + \cos^2(\theta_{T})$ distribution of the 
primary quarks. The direction of the 
$B$ candidate thrust axis for $B\bar{B}$ events is uniform.
The $B$ candidate direction, $\theta_{B}$, with respect 
to the beam axis  exhibits a $\sin^2(\theta_{B})$ distribution 
for $B\bar{B}$ events 
and is uniform for  $q\bar{q}$ events.

  A Fisher discriminant~\cite{Fisher} is formed from 
11 variables: the nine variables
of the ``Virtual Calorimeter'', $|\cos(\theta_{T})|$, and $|\cos(\theta_{B})|$. 
The discriminant $\cal{F}$ is the linear combination
\[ {\cal{F}} = \sum _{i=1}^{11}\alpha_i x_i \]
of the input variables, $x_i$, that maximizes the 
separation between signal and 
background. The coefficients $\alpha_i$ are determined from the Monte Carlo 
simulation using a large set of continuum events and signal events modeled 
as $B^+ \to K^+\pi^+\pi^-$.
We use the same set of coefficients $\alpha_i$ for all modes under study.
Figure~\ref{shape}b shows the ${\cal{F}}$ distributions for the Monte Carlo
signal in the mode $B^+\to K^+\pi^+\pi^-$, and the data signal in the mode
$B^+\to \bar{D^0}\pi^+$ followed by $\bar{D^0}\to K^+\pi^-$. 
The ${\cal{F}}$ distributions for Monte Carlo background and below-threshold 
background data for modes comprising three charged tracks are also presented
in Fig.~\ref{shape}b. The ${\cal{F}}$ distributions for both the signal and
background are fitted to Gaussian functions. 
The separation between the mean values of the signal and
background distributions is approximately 1.3 times the signal width.

  For the $K\pi\pi$ and $KK\pi$ final states we make 
the requirement on the Fisher discriminant 
variable ${\cal{F}}>0.8$; this rejects 90\% of continuum background
with about 54\% efficiency for the signal. In case of the $KKK$
final states, the continuum background is much smaller and we 
make the looser requirement
${\cal{F}}>0$. This rejects 53\% of continuum background with about
89\% efficiency for the signal.

  To determine the dominant sources of background from 
other $B$-meson decay modes we use a large set of Monte Carlo 
generated $B\bar{B}$ events where both $B$ mesons decay generically.
Most of the $B\bar{B}$ related background is found to originate 
from $B^+ \to \bar{D}^0\pi^+$ and $B^+ \to J/\psi(\psi(2S)) K^+$
decays. To suppress this type of background we apply the requirements
on invariant masses of two-particle combinations that are described below.
The background from the $B$ semileptonic decays is additionally 
suppressed by the electron veto requirement. 
The most significant background to $K^+\pi^+\pi^-$ final state from
the $B$ rare decays is found to originate from the $B^+\to \eta'K^+$
followed by $\eta'\to \rho^0\gamma$. We expect about 3\% of the
events of this type to satisfy all the selection criteria.
We find no significant background to the $K^+K^+K^-$ final state from 
other rare decays of $B$ mesons.

\section{Results of the Analysis }

\subsection{$B^+\to K^+\pi^+\pi^-$}

   For the $K^+\pi^+\pi^-$ final state, we select $B$ candidates formed 
from three charged tracks where one track is positively identified 
as a kaon  and the other two tracks are consistent with the pion hypothesis.
The resulting two dimensional $\Delta E$ versus $M_{BC}$ plot for all selected
$K^+\pi^+\pi^-$ combinations is presented in Fig.~\ref{kpp1_total}a where the
$B$ signal region is inside the box.
Figure~\ref{kpp1_total}b shows the Dalitz plot for candidates 
in the $B$ signal region. Large contributions from 
$B^+ \to \bar{D}^0\pi^+$ where $\bar{D}^0 \to K^+\pi^-$ 
and $B^+ \to J/\psi(\psi(2S)) K^+$ where $J/\psi(\psi(2S)) \to \mu^+\mu^-$ 
are apparent in the Dalitz plot. Modes with
$J/\psi(\psi (2S))$ contribute to this final state due to 
the muon-pion misidentification.
The contribution from the $J/\psi(\psi (2S)) \to e^+e^-$ submode 
is found to be negligible (less than 0.5\%) after the electron veto 
requirement. For further analysis we exclude $\bar{D}^0$ and 
$J/\psi (\psi(2S))$ signals by imposing requirements  
on the invariant masses of two intermediate particles:
\begin{center}
  $|M(K^+\pi^-)-1.865| > 0.100~~$GeV/$c^2$;  \\
\vspace*{0.3cm}
  $|M(h^+h^-)-3.097| > 0.070~~$GeV/$c^2$;~~~$|M(h^+h^-)-3.686| > 0.050~~$GeV/$c^2$,
\end{center}
where $h^+$ and $h^-$ are pion candidates.
For the $J/\psi(\psi(2S))$ rejection,
 we use the muon mass hypothesis for charged tracks 
to calculate $M(h^+h^-)$.

   The $\Delta E$ and $M_{BC}$ distributions 
for the events remaining after the exclusion of these large signals
are presented in Fig.~\ref{kpp1_mbde}a and
Fig.~\ref{kpp1_mbde}b, respectively.
Here a significant enhancement in the $B$ 
signal region is still observed;  the results of a fit to this
$\Delta E$ distribution are presented in Table~\ref{results1}.
The expected $\Delta E$ and $M_{BC}$ distributions,
which are the sum of luminosity-scaled off-resonance data and $B\bar{B}$ 
Monte Carlo, are shown
as open histograms in Figs.~\ref{kpp1_mbde}a and~\ref{kpp1_mbde}b,
respectively; the contributions from only the 
$B\bar{B}$  Monte Carlo sample are shown as hatched  histograms.
In the $\Delta E$ spectrum, the shape of the  $B\bar{B}$
background component is approximated as an exponential function with a 
parameter determined from the $B\bar{B}$ Monte Carlo.
As can be seen from the hatched histograms in Figs.~\ref{kpp1_mbde}a 
and~\ref{kpp1_mbde}b, there is no significant contribution to the 
signal from $B\bar{B}$ generic
decays after the large known backgrounds have been removed.

   To examine possible intermediate two-body states in the observed 
$B^+\to K^+\pi^+\pi^-$ signal,
we analyze the $K^+\pi^-$ and $\pi^+\pi^-$ invariant mass spectra shown in 
Figs.~\ref{kpp_hhmass}a and~\ref{kpp_hhmass}b, respectively.
To suppress the feed-across between the $\pi^+\pi^-$ and $K^+\pi^-$ states we 
require the $K^+\pi^-$ ($\pi^+\pi^-$) invariant mass to be larger than 
2.0(1.5)~GeV/c$^2$ when making the 
$\pi^+\pi^-$ ($K^+\pi^-$) projection. The hatched histograms  
shown in Figs.~\ref{kpp_hhmass}a and~\ref{kpp_hhmass}b are
the $h^+h^-$  invariant mass spectra 
for background events in the $\Delta E$ sidebands:
\begin{center}
 $ 5.272 < M_{BC} < 5.289~~$GeV/$c^2$~~~and\\
$-0.080 < \Delta E < -0.050~~$or$~~0.050 < \Delta E < 0.150~~$GeV,
\end{center}
scaled by area.

   The $K^+\pi^-$ invariant mass spectrum is characterized by a narrow peak 
around 0.9 GeV/$c^2$ which is identified as $K^{*0}(892)$
and a broad enhancement above 1.0 GeV/$c^2$ which is
subsequently referred to as $K_X(1400)$.

   In the $\pi^+\pi^-$ invariant mass spectrum two distinct 
structures in the low mass region are observed.
One is slightly below 1.0 GeV/$c^2$ which is identified as $f_0(980)$ 
and the other between 1.0 GeV/$c^2$ and 1.5 GeV/$c^2$ and referred to as 
$f_X(1300)$.
The peak around 3.4~GeV/$c^2$ is consistent with the process
$B^+ \to \chi_{c0}K^+$, $\chi_{c0} \to \pi^+\pi^-$, and is the subject
of a different analysis~\cite{chi_c0}. In this paper we exclude the $\chi_{c0}$
candidates from the analysis of two-body final states by applying the 
requirement on the $\pi^+\pi^-$ invariant mass:
$|M(\pi^+\pi^-)-3.415|>0.050$ GeV/$c^2$.

   For further analysis we subdivide the Dalitz plot 
area into seven non-overlapping
regions as defined in Table~\ref{kpp1_fits}. Regions from I to V are
arranged to contain the major part of the signal from the 
$B^+\to K^{*0}(892)\pi^+$, $B^+\to K_X(1400)\pi^+$,
$B^+\to \rho^0(770)K^+$, $B^+\to f_0(980)K^+$, and $B^+\to f_X(1300)K^+$
final states, respectively. The area in the Dalitz plot where 
$K\pi$ and $\pi\pi$ resonances overlap is covered by the 
region VI, and region VII covers the rest of the Dalitz 
plot. The $\Delta E$ and $M_{BC}$ distributions for each region are shown
in Fig.~\ref{pp_reso} and the results of the fits are summarized in
Table~\ref{kpp1_fits}. As can be seen from 
Fig.~\ref{pp_reso} and Table~\ref{kpp1_fits}, the contribution from
region VII to the total number of signal events is negligibly small.

\begin{table}[htb]
\caption{Results of the fit to the $\Delta E$ distribution for different 
         regions in the $K^+\pi^+\pi^-$ Dalitz plot. Columns list the definition
         of each region, reconstruction efficiency from Monte Carlo simulation,
         signal yield and statistical significance.}
\medskip
\label{kpp1_fits}
  \begin{tabular}{l c c c r}
   Region          & Mass range, GeV/$c^2$  &  Efficiency, \%        &
                   Yield, event    &  Significance, $\sigma$   \\ \hline
 I   & $0.82 < M(K\pi)   < 1.00$; $M(\pi\pi)>1.50$ 
                   & $20.7\pm3.7$ & $28.1^{+6.56}_{-5.92}$ & 6.1~~ \\
 II  & $1.00 < M(K\pi)   < 1.75$; $M(\pi\pi)>1.50$ 
                   & $19.2\pm1.3$ & $37.8^{+9.71}_{-9.05}$ & 4.7~~ \\
 III & $0.62 < M(\pi\pi) < 0.90$; $M(K\pi)>2.00$
                   & $16.7\pm2.4$ & $6.04^{+5.36}_{-4.68}$ & 1.3~~ \\
 IV  & $0.90 < M(\pi\pi) < 1.06$; $M(K\pi)>2.00$
                   & $19.9\pm3.2$ & $32.0^{+7.04}_{-6.36}$ & 6.9~~ \\
 V   & $1.06 < M(\pi\pi) < 1.50$; $M(K\pi)>2.00$ 
                   & $19.6\pm1.7$ & $25.4^{+7.54}_{-6.84}$ & 4.3~~ \\
 VI  & $M(K\pi)<2.00$; $M(\pi\pi) < 1.50$ 
                   & $14.7\pm3.3$ & $12.0^{+5.26}_{-4.67}$ & 2.9~~ \\
 VII & $M(K\pi)>2.00$; $M(\pi\pi) > 1.50$
                   & $16.1\pm0.5$ & $6.45^{+7.75}_{-7.05}$ & 0.9~~ \\ 
  \end{tabular}
\end{table}

\subsection{$B^+\to K^+K^+K^-$}

For the selection of $B\to K^+ K^+ K^-$ events,
we use events with three 
charged tracks that are positively identified as kaons. 
To suppress the background caused by $\pi/K$ misidentification, 
we exclude candidates if the invariant mass of any pair of oppositely 
charged tracks from the $B$ candidate 
is consistent with the $D^0\to K\pi$ hypothesis within 12 MeV 
($\sim 2\sigma$), independently of the PID information.

Figure~\ref{kkk_total}a 
shows the  two-dimensional $\Delta E$ versus $M_{BC}$ plot
for all selected $K^+K^+K^-$ combinations and
Fig.~\ref{kkk_total}b shows the
Dalitz plot for candidate events in the $B$ signal region.
Since in this case 
there are two same-charge kaons,
we distinguish the $K^+K^-$ combinations with smaller, $M(K^+K^-)_{min}$, 
and larger, $M(K^+K^-)_{max}$, invariant masses. 
We avoid double entries by forming the Dalitz plot as 
$M^2(K^+K^-)_{max}$ versus $M^2(K^+K^-)_{min}$,
as shown in Fig.~\ref{kkk_total}b.

  The signal from the Cabibbo-suppressed $B^+\to D^0_{CP}K^+$,
$D^0_{CP}\to K^+K^-$ decay mode is apparent as a vertical strip in
the Fig.~\ref{kkk_total}b Dalitz plot. The notation $D^0_{CP}$ means that
the $D$ meson decays to the CP eigenstate. 
The corresponding Cabibbo-allowed $B^+\to D^0_{CP}\pi^+$, 
$D^0_{CP}\to K^+K^-$ decays can 
also contribute to this final state as a result
of pion-kaon misidentification.
The detailed analysis of the decays of type $B^+\to D^0_{CP}K^+$
is described in ref.~\cite{dcpk}.

  We exclude candidates consistent with the
$B^+\to D^0_{CP}h^+$, $D^0_{CP}\to K^+K^-$
hypothesis from further analysis by imposing the requirement on the $K^+K^-$ 
invariant mass: 
\begin{center}
$|M(K^+K^-)-1.865|  > 0.025~~$GeV$/c^2$.
\end{center}
   The $\Delta E$ and $M_{BC}$ distributions
after the exclusion of $D$ mesons are presented in Figs.~\ref{kkk_mbde}a 
and~\ref{kkk_mbde}b, respectively. A large peak
in the $B$ signal region is apparent in both distributions.
The results of a fit to the $\Delta E$ distribution are presented in 
Table~\ref{results1}.

   The open histograms in Figs.~\ref{kkmass}a and~\ref{kkmass}b 
show the $M(K^+K^-)_{min}$ and $M(K^+K^-)_{max}$ distributions
for selected events, respectively; the hatched histograms
show the corresponding spectra for the $\Delta E$ sidebands:
\begin{center}
 $ 5.272 < M_{BC} < 5.289~~$GeV/$c^2$~~~and\\
$-0.200 < \Delta E <-0.050~~$or$~~0.050< \Delta E < 0.200~~$GeV,
\end{center}
scaled by area. 
The $M(K^+K^-)_{min}$ spectrum, Fig.~\ref{kkmass}a, is 
characterized by a narrow peak at 1.02 GeV/$c^2$ corresponding to the
$\phi(1020)$ meson and a broad structure around 1.5 GeV/$c^2$, which 
is subsequently referred to as $f_X(1500)$.
To exclude the possible contribution from the 
$B^+\to \chi_{c0}K^+$, $\chi_{c0}\to K^+K^-$ final state
we apply the requirement on the $K^+K^-$ invariant mass:
$|M(K^+K^-)-3.415|>0.050$ GeV/$c^2$. The study of this final 
state is described in ref.~\cite{chi_c0}.

  For further analysis we subdivide the area of the Dalitz 
plot into the four non-overlapping regions defined in Table~\ref{kkk_fits}.
Regions I and II are arranged to contain the major part of the signal 
from the $B^+\to \phi(1020)K^+$ and $B^+\to f_X(1500)K^+$
final states respectively. Regions III and IV cover the rest part of the 
Dalitz plot.
The $\Delta E$ and $M_{BC}$ distributions for each region are shown
in Fig.~\ref{kk_reso} and the results of the fit are summarized in
Table~\ref{kkk_fits}.

\begin{table}[htb]
\caption{Results of the fit to the $\Delta E$ distribution
         for different regions in the $K^+K^+K^-$ Dalitz plot.
         Columns list the definition
         of each region, reconstruction efficiency from Monte Carlo simulation,
         signal yield and statistical significance.}
\medskip
\label{kkk_fits}
  \begin{tabular}{l c c c r}
  Region   &  Mass range, GeV/$c^2$     & Efficiency,   \%   &
                  Yield, event & Significance, $\sigma$             \\ \hline
 I   & $1.005<M(KK)_{min}<1.035$ 
               & $24.6\pm2.5$ & $24.6^{+5.87}_{-5.23}$ & 6.7~~ \\
 II  & $1.035<M(KK)_{min}<2.00$
               & $23.3\pm0.8$ & $84.0^{+12.5}_{-11.8}$ & 8.9~~ \\
 III & $M(KK)_{min}>2.00$; $M(KK)_{max}>3.40$
               & $23.9\pm1.1$ & $15.9^{+6.26}_{-5.55}$ & 3.3~~ \\
 IV  & $M(KK)_{min}>2.00$; $M(KK)_{max}<3.40$
               & $24.7\pm0.8$ & $26.1^{+6.11}_{-5.43}$ & 6.7~~ \\
  \end{tabular}
\end{table}

\section{Branching Fractions Results}

   To determine branching fractions we 
normalize our results to the observed 
$B^+\to \bar{D}^0\pi$, $\bar{D}^0\to K^+\pi^-$ signal. 
Although this introduces a $9.7$\% systematic error 
due to the uncertainty in the
$B^+\to \bar{D}^0\pi$ branching fraction, it removes systematic
effects in the particle identification efficiency,
charged track reconstruction efficiency and the 
systematic uncertainty due to the cuts on event shape variables.

   We calculate the branching fraction for $B$ meson decay to a 
particular final state $f$ via the relation:
\begin{equation}
   {\cal{B}}(B^+\to f) = 
   {\cal{B}}(B^+\to \bar{D}^0\pi^+)\times{\cal{B}}(\bar{D}^0\to K^+\pi^-)
   \frac{N_f}{N_{D\pi}}\times\frac{\varepsilon_{D\pi}}{\varepsilon_{f}},
\end{equation}
where $N_f$ and $N_{D\pi}$ are the numbers of 
reconstructed events for the particular final state 
$f$ and for the reference process, respectively;
$\varepsilon_{f}$ and $\varepsilon_{D\pi}$ are the corresponding 
reconstruction efficiencies.

   The number of signal events for the normalization processes
$B^+\to \bar{D}^0\pi$ and $\bar{D}^0\to K^+\pi^-$ is found
to be $1137\pm38$ for the
$K^+K^+K^-$ selection requirements and $619\pm29$ for  
requirements used for the other $Khh$ combinations with the
reconstruction efficiencies of 31.3\% and 16.8\% respectively.

%%%%%%%%%%%%%%%%%%%%%%%%%%%%%%%%%%%%%%%%%%%%%%%%%%%%%%%%%%%%%%%%%%%%%%

\subsection{Three-body branching fractions}

   For branching fraction calculations
we use the signal yield extracted from the fit to the 
corresponding $\Delta E$ distribution since the $M_{BC}$ distribution
suffers more from $B\bar{B}$ background.
   The reconstruction efficiency for three-body final states is determined
from the Monte Carlo simulations of events that are generated to have a 
uniform distribution over the Dalitz plot.

 The branching fraction results are 
summarized in Table~\ref{results1}. The first quoted error is statistical
and the second is systematic.
We do not observe a statistically significant signal in 
the $K^-\pi^+\pi^+$,
$K^+K^+\pi^-$ or $K^+K^-\pi^+$ final states and place 90\% confidence
level upper limits on their respective branching fractions.

\begin{table}[htb]
\caption{Measurement results. Branching fractions and 90\% C.L. upper limits 
         for $B^+\to K^+h^+h^-$ final states.}
\medskip
\label{results1}
  \begin{tabular}{l c c c} 
    Mode              &     Efficiency, \%    &
                            Yield, event      &
                             ${\cal{B}}, 10^{-6}$  \\ \hline
  $K^+\pi^-\pi^+$  & 16.6 &     $177\pm20$         & $58.5\pm7.1\pm8.8$ \\
  $K^+K^+K^-$      & 24.5 &     $162\pm16$         & $37.0\pm3.9\pm4.4$ \\
  $K^-\pi^+\pi^+$  & 16.2 & $3.86^{+8.23}_{-7.75}$ & $<7.7$      \\
  $K^+K^+\pi^-$    & 13.3 & $6.78^{+4.87}_{-4.17}$ & $<6.0$      \\
  $K^+K^-\pi^+$    & 13.2 & $32.9^{+9.49}_{-8.80}$ & $<21$      \\
  \end{tabular}
\end{table}

   The following sources of systematic errors are found to be the dominant: 
\begin{itemize}
  \item{ uncertainty due to the nonuniformity of the reconstruction efficiency 
         over the Dalitz plot. It is found to be 7.6\% for $K^+\pi^-\pi^+$
         and 3.7\% for $K^+K^+K^-$;}
  \item{ uncertainty in $B^+\to\bar{D}^0\pi^+$ and $\bar{D}^0\to K^+\pi^-$ 
         branching fractions: 9.7\%;}
  \item{ uncertainty in the parameterization of the signal and background 
         shape in $\Delta E$. This is 8.5\% for $K^+\pi^-\pi^+$
         and 4.7\% for $K^+K^+K^-$;}
  \item{ kaon identification efficiency for modes with more than one 
         kaon in the final state: 3\% per kaon; }
\end{itemize}

%%%%%%%%%%%%%%%%%%%%%%%%%%%%%%%%%%%%%%%%%%%%%%%%%%%%%%%%%%%%%%%%%%%%%%

\subsection{Exclusive two-body branching fractions 
in the $K^+\pi^+\pi^-$ final state.}

   In the determination of the branching fractions for exclusive 
two-body final states,
we take into account the possibility of interference between 
wide resonances.
This requires some assumptions about the states that are being observed
and, as a consequence, introduces some model dependence
into the extraction of the exclusive 
branching fractions.  The present level of statistics does
not permit unambiguous interpretation of the $K_X(1400)$ and $f_X(1300)$
states and, thus,  it is not possible to
use the data to fix all of the input model parameters.
For this analysis we assume that the observed $K_X(1400)$ and $f_X(1300)$
states are $0^{++}$ scalars.  While this does not contradict
the observed signal,  some contributions from
vector ($1^{-}$) and tensor ($2^{+}$) resonances can not be 
excluded. The uncertainty related to this assumption is included 
in the model-dependent error
described below. We ascribe to the $K_X(1400)$ state the parameters of 
$K^*_0(1430)$ ($M$ = 1412 MeV/$c^2$, $\Gamma$ = 294 MeV)
and to $f_X(1300)$ state the parameters of 
$f_0(1370)$ ($M$ = 1370 MeV/$c^2$, $\Gamma$ = 400 MeV)~\cite{PDG}.

   For further analysis we make following assumptions:
\begin{itemize}
  \item{ The observed signal in the $K^+\pi^+\pi^-$ final state can be 
         described by some number of two-body final states. 
         We restrict ourselves with the following set of exclusive 
         two-body final states: $K^{*0}(892)\pi^+$, $K_X(1400)\pi^+$,
         $\rho^0(770)K^+$, $f_0(980)K^+$ and $f_X(1300)K^+$. 
         We enumerate these final states as 1 through 5 in the order they 
         are mentioned above.}

  \item{ Given this set of two-body final states, we determine the 
         exclusive branching fractions neglecting the effects of 
         interference. The uncertainty due to possible interference
         between different intermediate states is included 
         in the final result as a model-dependent error.}
\end{itemize}

   In order to extract the signal yield for each two-body final 
state, we perform a simultaneous likelihood
fit to the $\Delta E$ distributions
for the seven regions of the $K^+\pi^+\pi^-$ Dalitz plot
(see Fig.~\ref{pp_reso} and Table~\ref{kpp1_fits}).
We express the expected number $n_j$ of signal events 
in the $j$-th region of the Dalitz plot as a linear combination:
\[ n_j = \sum^{5}_{i=1}\varepsilon_{ij}N_i,\]
where $N_i$ is the total number of signal events in the $i$-th 
two-body final state and $\varepsilon_{ij}$ is the probability
for the $i$-th final state to contribute to the $j$-th region in
the Dalitz plot. The $\varepsilon_{ij}$ matrix is determined 
from a Monte Carlo simulation and includes the 
reconstruction efficiency.
This procedure takes correlations 
between different channels into account when
determining the statistical
errors.

   The results of the fit are summarized in Table~\ref{kpp_min}.
Combining all the relevant numbers and using Eq.~1,
we calculate the product of branching fractions 
${\cal{B}}(B^+\to Rh^+)\times{\cal{B}}(R\to h^+h^-)$,
where $R$ denotes the two-body intermediate resonant state.
The branching fraction result for the $B^+\to K^{*0}(892)\pi^+$
final state is in agreement with the results of a separate study
of $B$ meson decays to the pseudo-scalar and vector 
final states~\cite{b2kstp}.

  We present three types of error for the branching fractions: 
the first error is
statistical, the second is systematic, and the third
reflects the model-dependent uncertainty.   In general, the 
model-dependent error is due to uncertainties in the effects of interference
between different resonant states. We estimate this error by means of
a $B^+\to K^+\pi^+\pi^-$ Monte Carlo simulation that includes 
interference effects between all final states mentioned above.
We vary the relative phases of resonances 
and determine the signal yield using the procedure described above.
The maximal deviations from the central values are used as
an estimate of the model dependence of the obtained branching 
fractions.

\begin{table}[htb]
\caption{Results of the simultaneous fit to the $K^+\pi^+\pi^-$ final state.}
\medskip
\label{kpp_min}
  \begin{tabular}{lcccc} 
    Two body mode   & Efficiency, \%         & Yield, events 
                    & Significance, $\sigma$ 
    & ${\cal{B}}_{B^+\to Rh^+}\times{\cal{B}}_{R\to h^+h^-}$, $10^{-6}$  \\ \hline 
 $K^{*0}(892)\pi^+$ & 19.1 & $38.5^{+8.50}_{-7.90}$ & 6.2
                    & $11.1^{+2.5+1.4+2.0}_{-2.3-1.4-3.9}$ \\
  $K_X(1400)\pi^+$  & 17.0 & $39.1^{+10.8}_{-10.5}$ & 4.1 
                    & $12.7^{+3.5+1.8+2.9}_{-3.4-1.8-5.8}$  \\
 $\rho^0(770)K^+$   & 18.9 & $1.75^{+8.60}_{-7.38}$ & 0.2 
                    & $<9.6$  \\
   $f_0(980)K^+$    & 19.3 & $40.9^{+8.80}_{-9.56}$ & 6.0 
                    & $11.7^{+2.5+1.5+4.1}_{-2.7-1.5-1.0}$  \\
  $f_X(1300)K^+$    & 17.3 & $33.6^{+12.1}_{-11.1}$ & 3.2 
                    & $10.7^{+3.9+1.4+6.9}_{-3.5-1.4-2.8}$  \\
  \end{tabular}
\end{table}

   We find that the model-dependent errors associated
with the wide resonances introduce 
significant uncertainties into the branching fraction determination. In the 
case of the $\rho^0(770)K^+$ final state, this effect is enhanced by the
smallness of the signal itself. The effects of interference of a small 
$\rho^0(770)$ signal with the large $f_0(980)$  signal or
with a broad $f_X(1300)$ resonance could result in as much as a doubling
of the observed $\rho^0(770)K^+$ final state or its
total suppression.  Since we do not observe a significant signal in
this channel, we report a 90\% confidence level upper limit. 

%%%%%%%%%%%%%%%%%%%%%%%%%%%%%%%%%%%%%%%%%%%%%%%%%%%%%%%%%%%%%%%%%%%%%%

\subsection{Exclusive two-body branching fractions in 
$K^+K^+K^-$ final state.}

   In the case of the three charged kaons final state we clearly
observe the presence of the $\phi(1020)$ meson plus a very broad
$f_X(1500)$ structure that we currently cannot interpret
unambiguously. It could be a complex superposition of
several intermediate states and some contribution from
non-resonant $B^+\to K^+K^+K^-$ decay is also possible. For our 
study of systematic and model-dependent uncertainties, we construct 
a simplified model and parameterize the $f_X(1500)$ structure
as a hypothetical scalar state with $M=1500$ MeV/$c^2$ and
$\Gamma = 700$ MeV. We find qualitative agreement between the
experimental Dalitz plot distribution of the signal events and
that obtained from the Monte Carlo simulation with this simple model.

  Then we extract the signal yield for the two-body final states:
$B^+\to\phi(1020)K^+$ and the so-called $B^+\to f_X(1500)K^+$, which
is, in fact, all of the remaining signal.
We follow the same procedure as we used for the $K^+\pi^+\pi^-$ final state.
The signal yields are determined from a simultaneous fit to 
the $\Delta E$ distributions for four separate regions of the $K^+K^+K^-$ 
Dalitz plot (see Fig.~\ref{kk_reso} and Table~\ref{kkk_fits}).
The results of the fit are summarized in Table~\ref{kkk_min}.
The branching fraction result for the $B^+\to\phi(1020)K^+$
final state is in good agreement with the results of a dedicated
analysis of the $B^+\to\phi(1020)K^+$ and $B^+\to\phi(1020)K^{*+}(892)$
final states~\cite{b2phik}, 
${\cal{B}}(B^+\to\phi(1020)K^+) = (10.6^{+2.1}_{-1.9}\pm2.2)\times10^{-6}$. 
This latter number should be considered as the
current ``official'' Belle result for
the $B^+\to\phi(1020)K^+$ branching fraction.

\begin{table}[htb]
\caption{Results of the simultaneous fit for the $K^+K^+K^-$ final state.}
\medskip
\label{kkk_min}
  \begin{tabular}{lcccc} 
    Two body mode   & Efficiency, \%         & Yield, events 
                    & Significance, $\sigma$ 
  & ${\cal{B}}_{B^+\to RK^+}\times{\cal{B}}_{R\to K^+K^-}$, $10^{-6}$  \\ \hline 
  $\phi(1020)K^+$   & 25.1 & $30.1^{+7.35}_{-6.52}$ & 6.4
                    & $6.70^{+1.6+0.8+0.4}_{-1.5-0.8-0.4}$ \\
  $f_X(1500)K^+$    & 22.1 & $122^{+15.1}_{-14.5}$ & 12.1 
                    & $30.8^{+3.8+3.9+1.5}_{-3.7-3.9-1.5}$ \\
  \end{tabular}
\end{table}

  We determine the model-dependent error the same way as we did for
the $K^+\pi^+\pi^-$ final state. In the case of the $K^+K^+K^-$ final
state the  model dependent error is found to be much smaller then
in $K^+\pi^+\pi^-$ final state. This is mainly due to the
narrow width of the $\phi(1020)$ meson.

\section{Discussion \& Conclusion}

 The high quality Belle's $\pi/K$ separation allows us to measure for the 
first time the branching ratios for the three-body modes 
$ {\cal{B}}(B^+\to K^+\pi^-\pi^+)~~= (58.5\pm7.1\pm 8.8)\times10^{-6}$ and 
$ {\cal{B}}(B^+\to K^+K^-K^+)     = (37.0\pm3.9\pm 4.4)\times10^{-6}$
without assumptions about particular intermediate mechanisms.
CLEO~\cite{berg96b} and BaBar~\cite{babar} have previously placed upper
limits on non-resonant three-body decays;
the reported numbers for $B^+ \to K^+\pi^+\pi^-$ 
(CLEO: ${\cal{B}}(B^+ \to K^+\pi^+\pi^-)<28 \times 10^{-6}$, 
BaBar: ${\cal{B}}(B^+ \to K^+\pi^+\pi^-)<66 \times 10^{-6}$)
and $B^+ \to K^+K^+K^-$ 
(CLEO: ${\cal{B}}(B^+ \to K^+K^+K^-)<38 \times 10^{-6}$) are 
lower than those presented in this paper.
A comparison of the applied selection criteria 
shows that CLEO and BaBar restricted their analyses to the
region of the invariant masses above 2~GeV/$c^2$ for any pair of the
particles. This requirement effectively removes most of the low mass
resonances that provide the dominant contribution to our observed signal.
They assume a uniform distribution of events over the Dalitz plot
to obtain the limits quoted above.

   The upper limits reported here for the $K^-\pi^+\pi^+$, $K^+K^-\pi^+$ 
and $K^+K^+\pi^-$ modes are considerably more restrictive 
than previous limits from
CLEO~\cite{berg96b}
(${\cal{B}}(B^+ \to K^-\pi^+\pi^+)<56 \times 10^{-6}$,
${\cal{B}}(B^+ \to K^+K^-\pi^+)<75 \times 10^{-6}$)
and OPAL~\cite{OPAL}
(${\cal{B}}(B^+ \to K^+K^+\pi^-)<87.9 \times 10^{-6}$). 
%A search for the $K^+K^+\pi^-$ mode 
%is of particular interest since in some extensions of the Standard Model 
%this branching fraction is predicted to be significantly enhanced.

   Significant signals are observed for the first time in the decay modes
$B^+ \to K^{*0}(892) \pi^+$ and $B^+ \to f_0(980) K^+$ with  
branching fraction products of 
${\cal{B}}(B^+\to K^{*0}(892)\pi^+)\times{\cal{B}}(K^{*0}(892)\to K^+\pi^-) =
(11.1^{+2.5+1.4+2.0}_{-2.3-1.4-3.9})\times10^{-6}$ and
${\cal{B}}(B^+\to f_0(980)K^+)\times{\cal{B}}(f_0(980)\to \pi^+\pi^-) =
(11.7^{+2.5+1.5+4.1}_{-2.7-1.5-1.0})\times10^{-6}$, 
respectively. The latter final state is of interest since this
is the first observation of a $B$ decay to a charmless 
scalar-pseudoscalar final state. The significant enhancement 
in the $K^+\pi^-$ invariant mass spectrum above the 
$K^*(892)$ mass (referred to as $K_X(1400)$) 
agrees with the scalar $K_0^*(1430)$ hypothesis. This is also
in agreement with theoretical predictions~\cite{chernyak,maltman} for the 
$B^+\to K_0^*(1430)\pi^+$ branching fraction made based on the factorization 
model. Nevertheless, we cannot exclude some contribution from the
tensor $K_2^*(1430)$ state.

  Large uncertainties arise in the interpretation of the peak with a 
$\pi^+\pi^-$ invariant mass around 1300 MeV/$c^2$ in the $K\pi\pi$ system. 
There are two known candidate states: the $f_2(1270)$ and $f_0(1370)$~\cite{PDG}.
Attributing the peak to the $f_0(1370)$, 
with its rather small coupling to $\pi^+\pi^-$~\cite{anisov}, 
would lead to an unusually large branching fraction 
for a charmless $B$ decay mode.
On the other hand, as recently shown in~\cite{kim}, 
factorization model predicts a very small
branching ratio for $B^+\to f_2(1270)K^+$.
If our observation is, in fact, due to the $f_2(1270)$, 
this would provide evidence for a significant nonfactorizable contribution. 

  We cannot identify the broad structure observed in the $B^+ \to K^+K^+K^-$ 
final state above $\phi(1020)$ meson.
It is hardly compatible with the presence of single scalar state either 
$f_0(1370)$ and $f_0(1500)$~\cite{PDG}.
We also cannot exclude the presence of a non-resonant contribution
or the case of several
resonances contributing to the excess 
in the $K^+K^-$ invariant mass spectrum seen around 1.5 GeV/$c^2$.

We find that effects of interference between 
different two-body intermediate states can have a significant 
influence on the observed two-particle mass spectra and a
full amplitude analysis of three-body $B$ meson decays is required
for a more complete understanding.
This will be possible with increased statistics.

\section*{Acknowledgement}

We wish to thank the KEKB accelerator group for the excellent
operation of the KEKB accelerator. We acknowledge support from the
Ministry of Education, Culture, Sports, Science, and Technology of Japan
and the Japan Society for the Promotion of Science; the Australian
Research
Council and the Australian Department of Industry, Science and
Resources; the Department of Science and Technology of India; the BK21
program of the Ministry of Education of Korea and the CHEP SRC
program of the Korea Science and Engineering Foundation; the Polish
State Committee for Scientific Research under contract No.2P03B 17017;
the Ministry of Science and Technology of Russian Federation; the
National Science Council and the Ministry of Education of Taiwan; the
Japan-Taiwan Cooperative Program of the Interchange Association; and
the U.S. Department of Energy.

\newpage

\begin{figure}[t]
\centering
  \includegraphics[height=10.0cm,width=12.0cm]{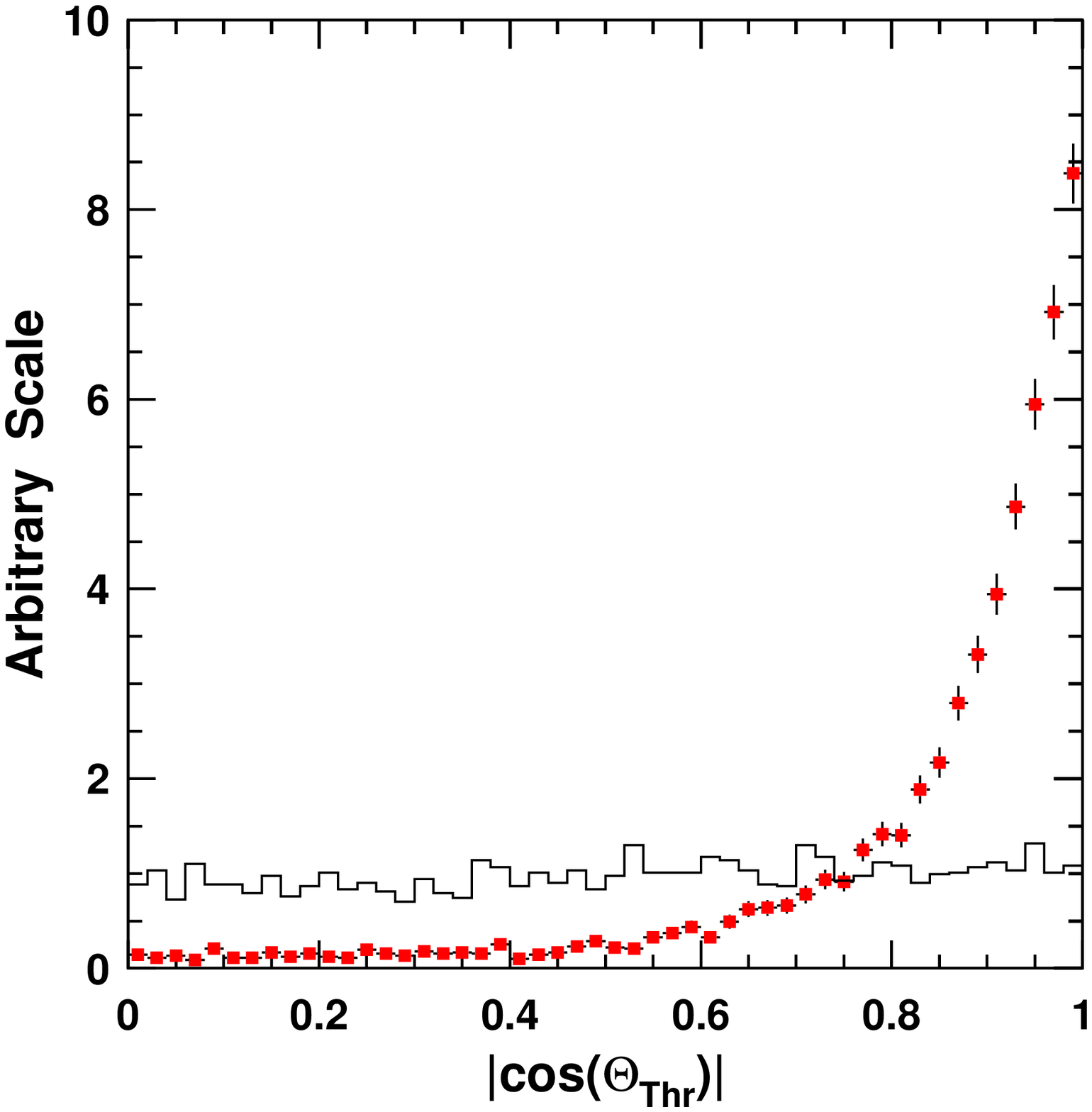} \\ %\hfill
  \includegraphics[height=10.0cm,width=12.0cm]{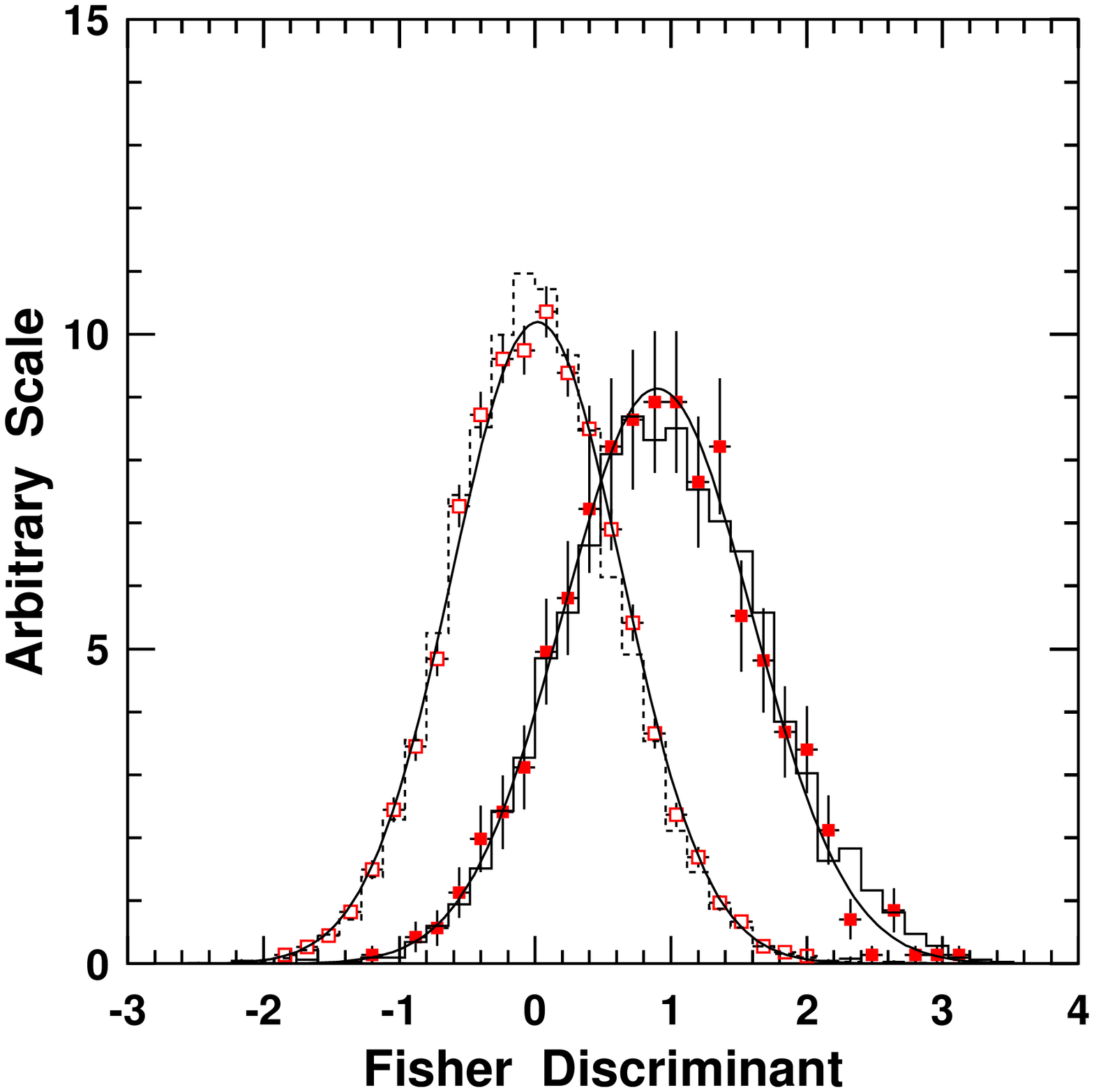}
  \centering

  \caption{{\bf a)} The $|cos(\theta_{Thr})|$ distribution for 
           $B^+\to K^+\pi^+\pi^-$ signal Monte Carlo events 
           (solid histogram) and off-resonance data (filled squares).
           {\bf b)} The ${\cal{F}}$ distribution for $B^+\to K^+\pi^+\pi^-$ 
           Monte Carlo (solid histogram), $B^+\to \bar{D^o}\pi^+$
           signal data (filled squares), continuum Monte Carlo 
           (dashed histogram), off-resonance data (open squares). 
           The curves are fits to the data.}
  \label{shape}
\end{figure}

%%%%%%%%%%%%%%%%%%%%%%%%%%%%%%%%   Plots for Kpipi section   %%%%%%%%%%%%%%%%%%%%%%%%%%

\begin{figure}[hbt]
  \centering
  \includegraphics[height=10.0cm,width=12.0cm]{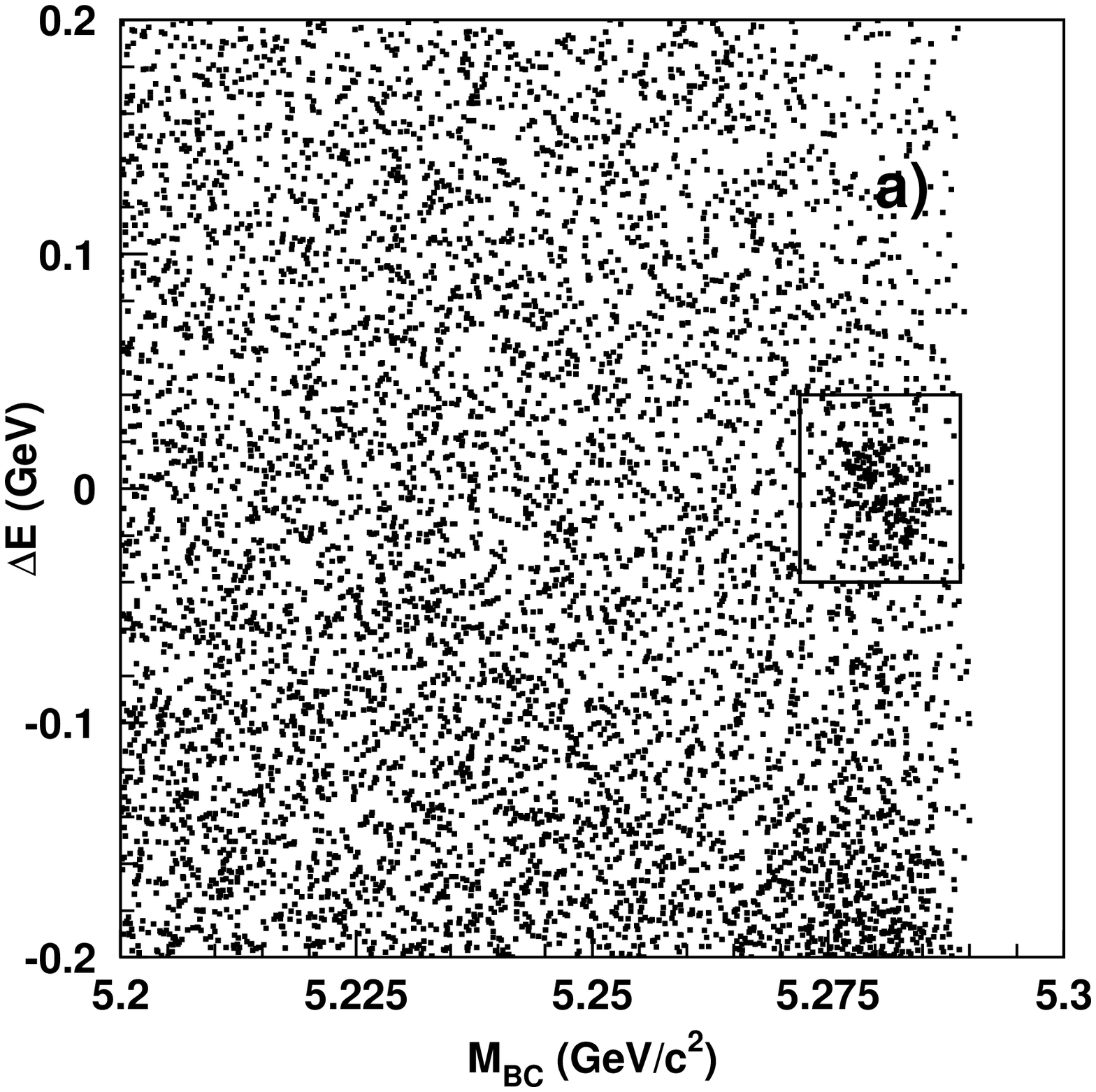} \hfill
  \includegraphics[height=10.0cm,width=12.0cm]{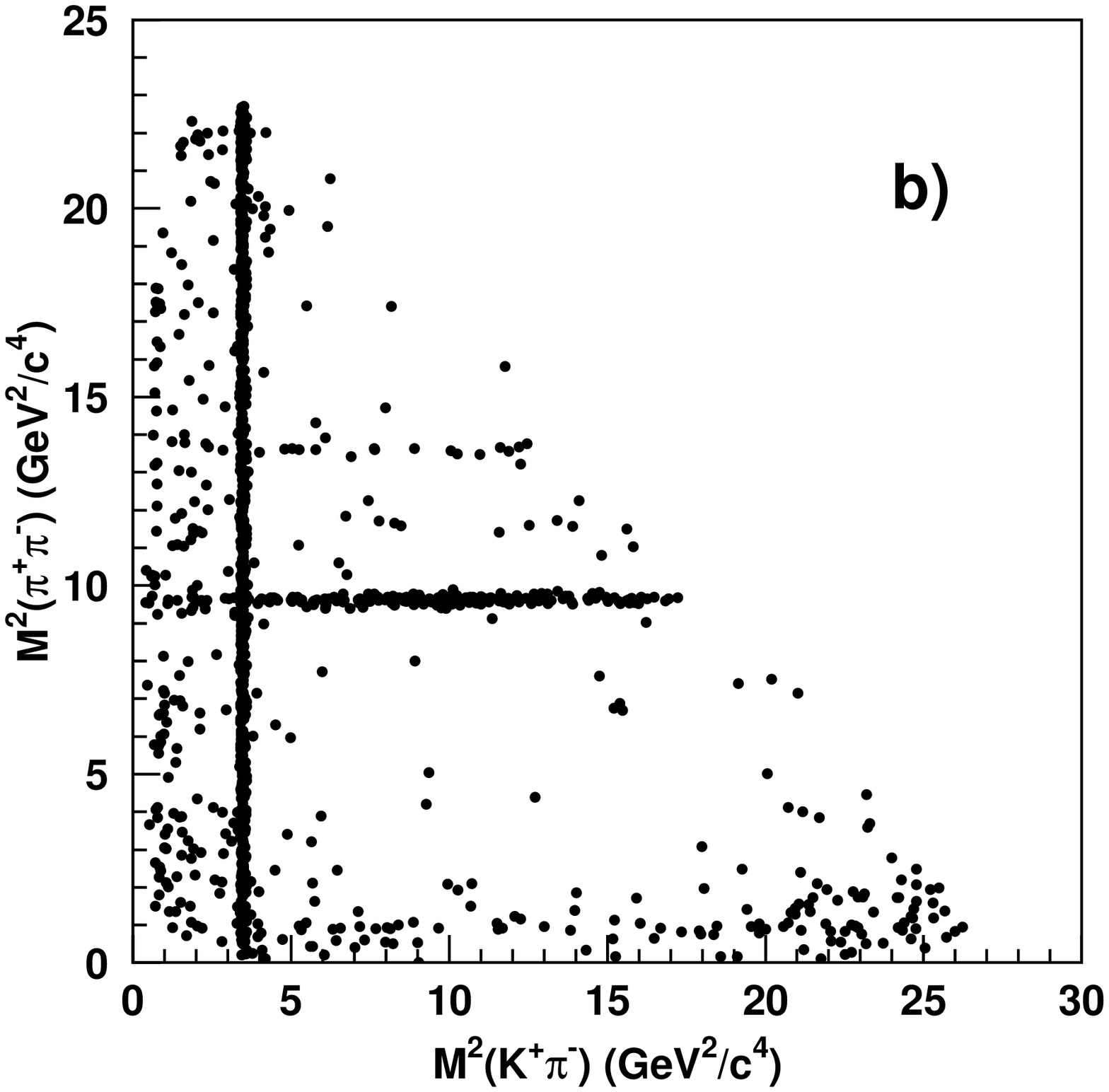}
  \centering
 
  \caption{ {\bf a)} The $\Delta E$ versus $M_{BC}$ plot for all 
                     $B^+ \to K^+\pi^-\pi^+$ candidates.
            {\bf b)} The Dalitz Plot for $B^+ \to K^+\pi^-\pi^+$ candidates
                     from the $B$ signal region inside the box in a).}
  \label{kpp1_total}
\end{figure}
\begin{figure}[htb]
  \includegraphics[height=8.0cm,width=8.0cm]{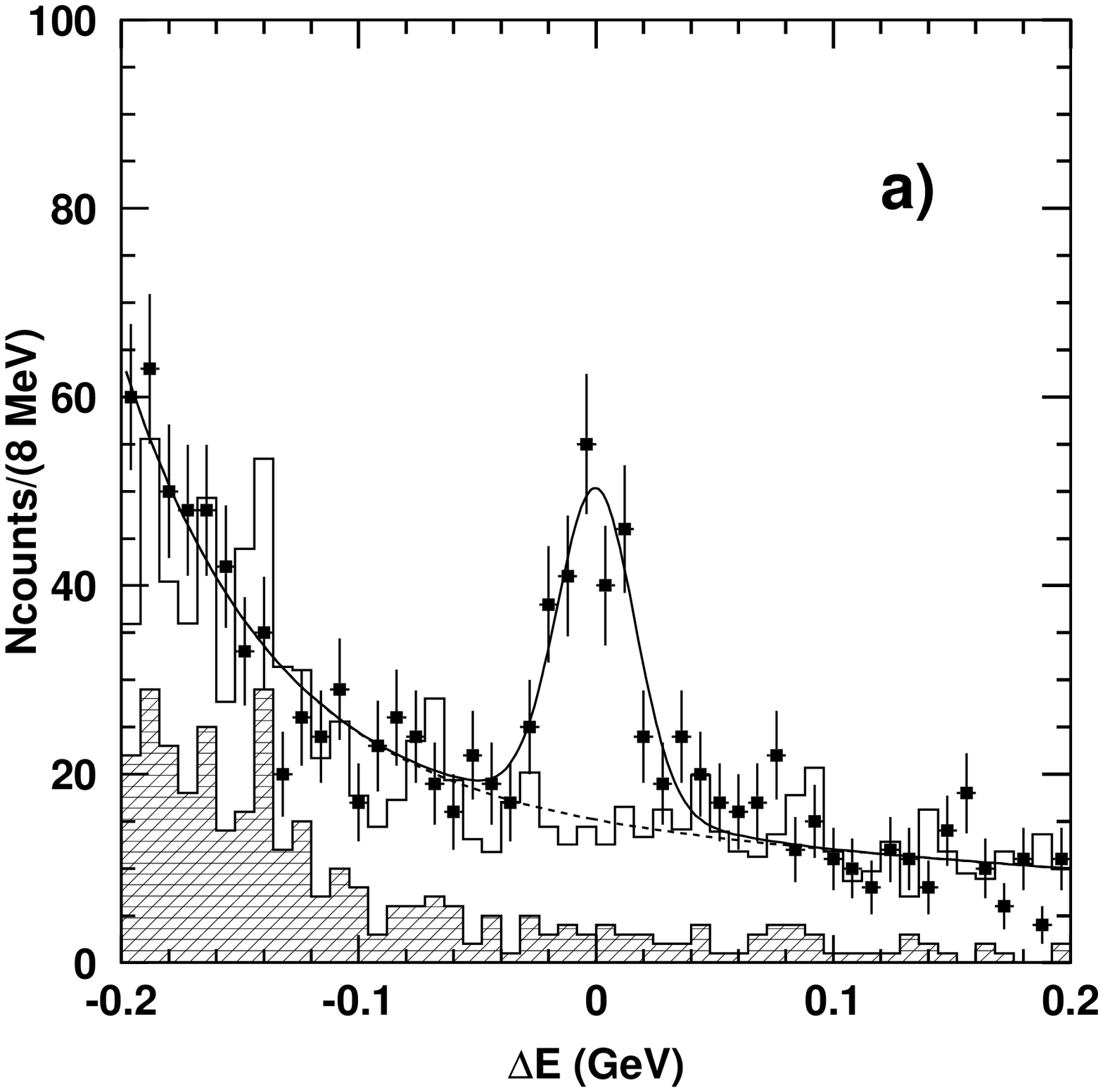} \hfill
  \includegraphics[height=8.0cm,width=8.0cm]{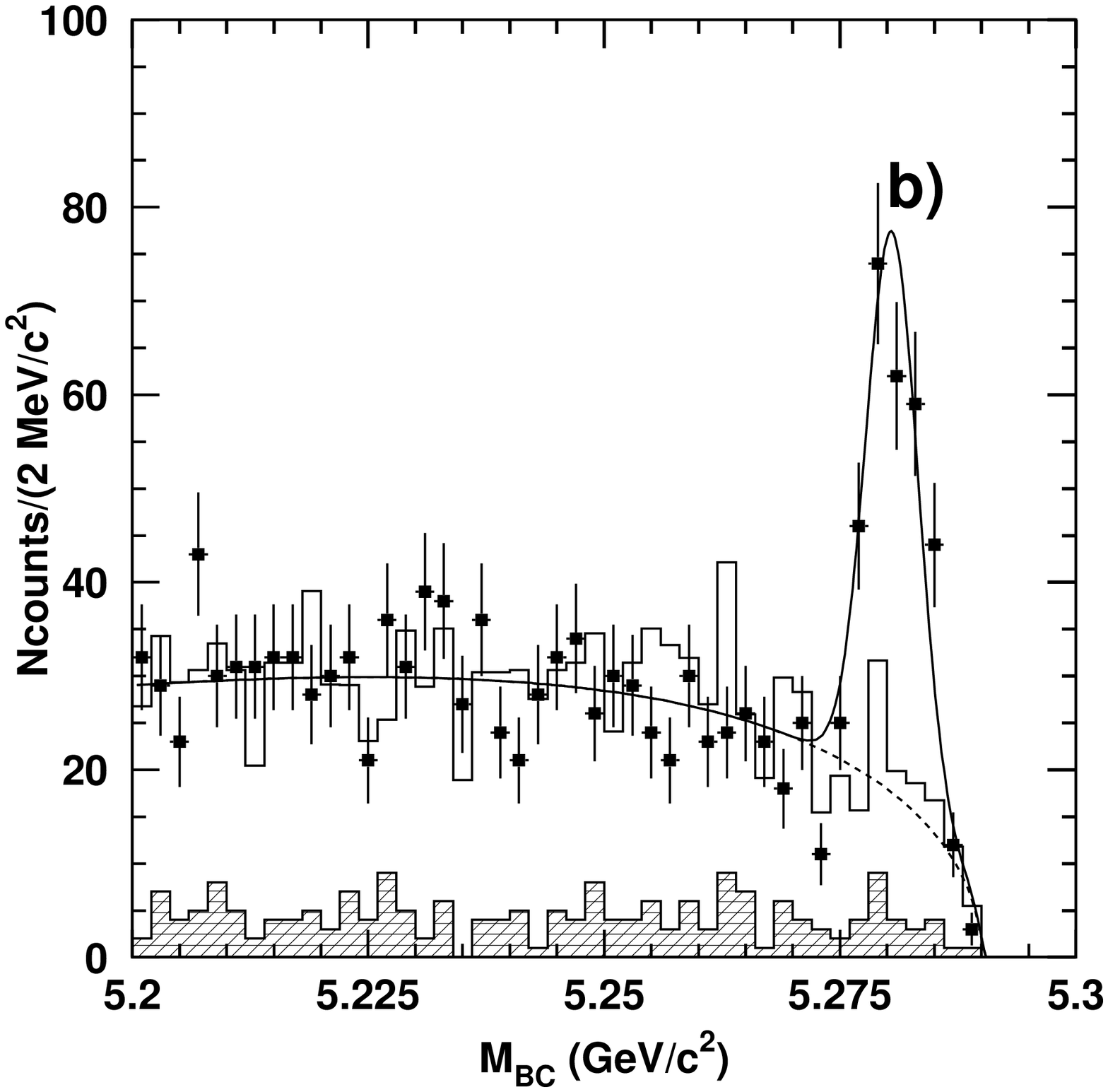}
  \caption{The $\Delta E$ - a) and $M_{BC}$ - b) distributions for the
           $B^+ \to K^+\pi^-\pi^+$ final state.
           Candidates consistent with $B^+ \to \bar{D}^0\pi^+$ or 
           $B^+ \to J/\psi(\psi') K^+$ are excluded. Points are data, 
           open histograms are the proper sum of the off-resonance data 
           and $B\bar{B}$ Monte Carlo, and hatched histograms show 
           the contribution of 
           $B\bar{B}$ Monte Carlo only. The curves show the fit to the data.}
  \label{kpp1_mbde}
\end{figure}

\begin{figure}[htb]
  \includegraphics[height=8.0cm,width=8.0cm]{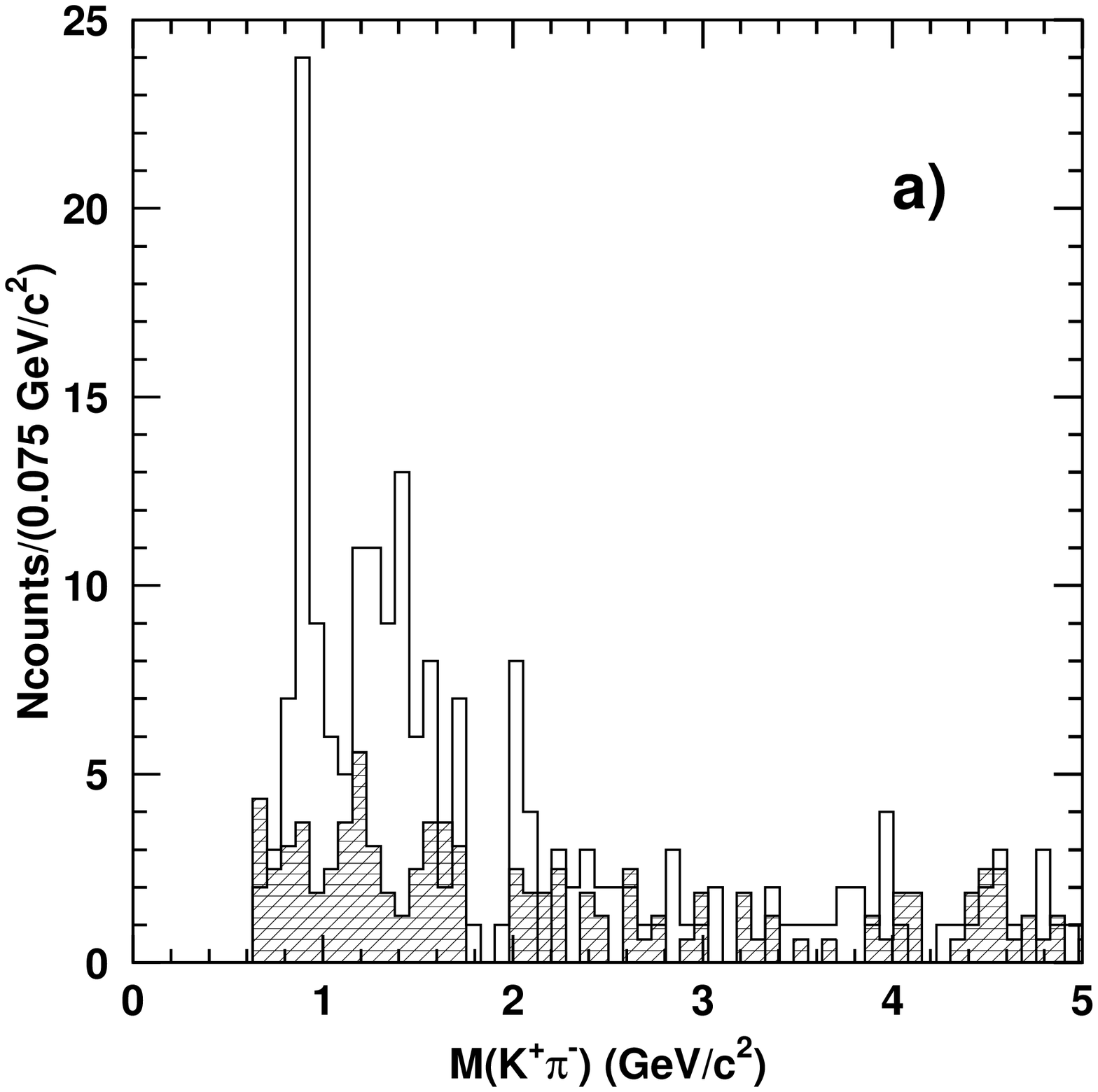} \hfill
  \includegraphics[height=8.0cm,width=8.0cm]{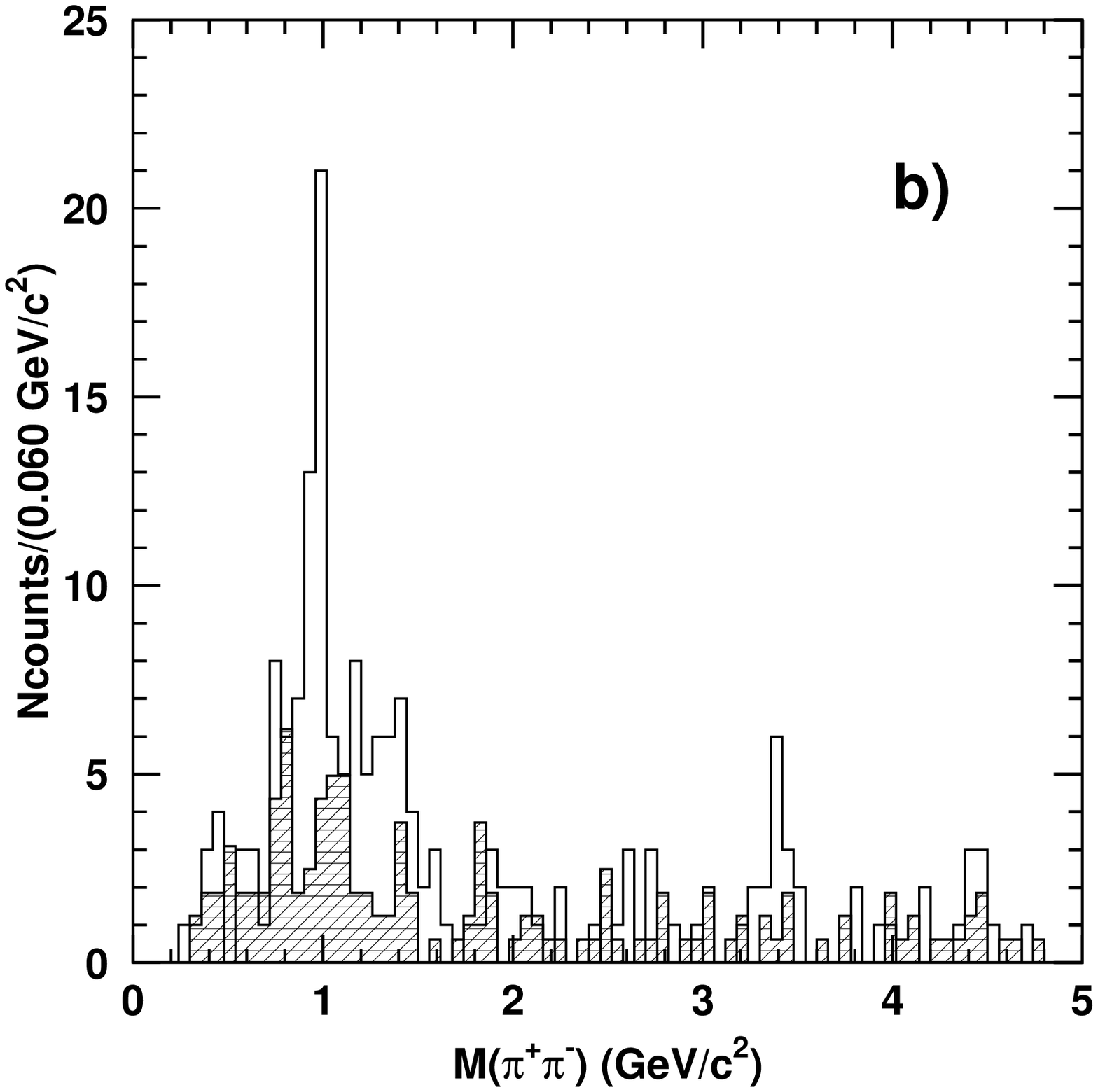} \\
  \caption{ The $K^+\pi^-$ - a) and $\pi^+\pi^-$ - b) invariant mass spectra
            for $B^+\to K^+\pi^-\pi^+$ candidates. Open histogram for candidates 
            from the $B$ signal region, hatched histogram for candidates 
            from the $\Delta E$ sidebands. See the text for details.}
  \label{kpp_hhmass}
\end{figure}

\begin{figure}[htb]
    \includegraphics[height=3.0cm,width=8.0cm]{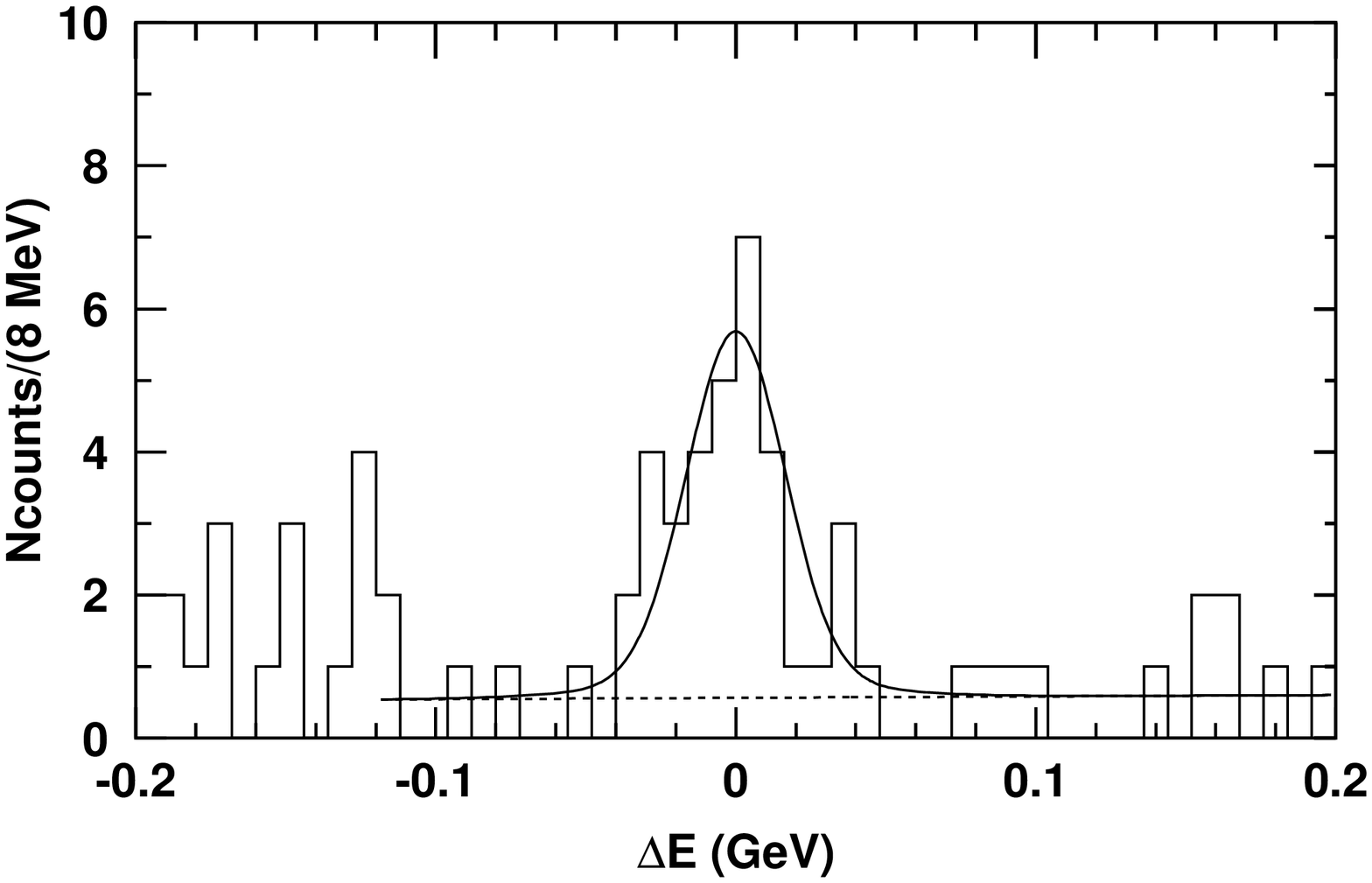} \hfill
    \includegraphics[height=3.0cm,width=8.0cm]{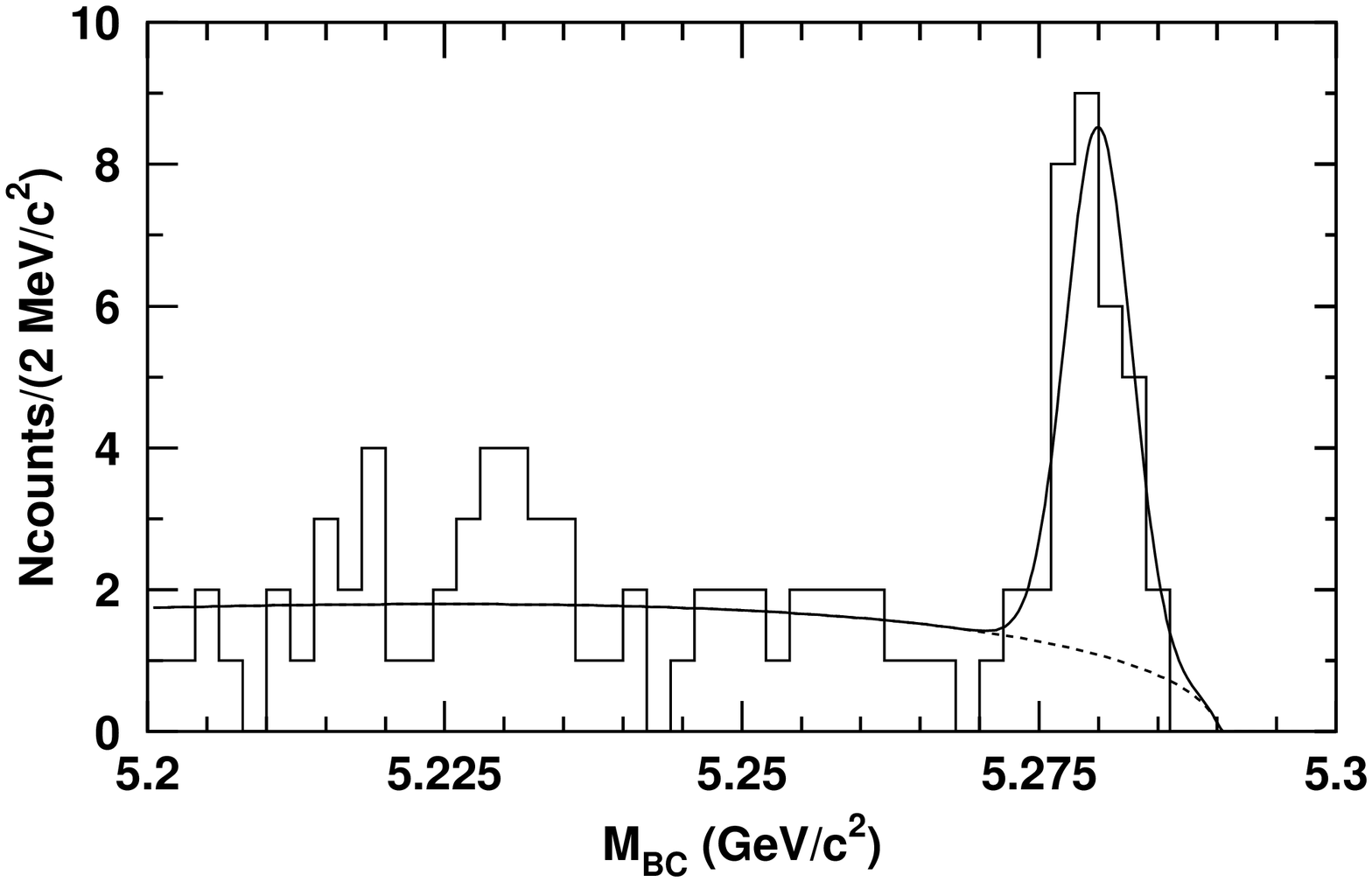} \\
    \includegraphics[height=3.0cm,width=8.0cm]{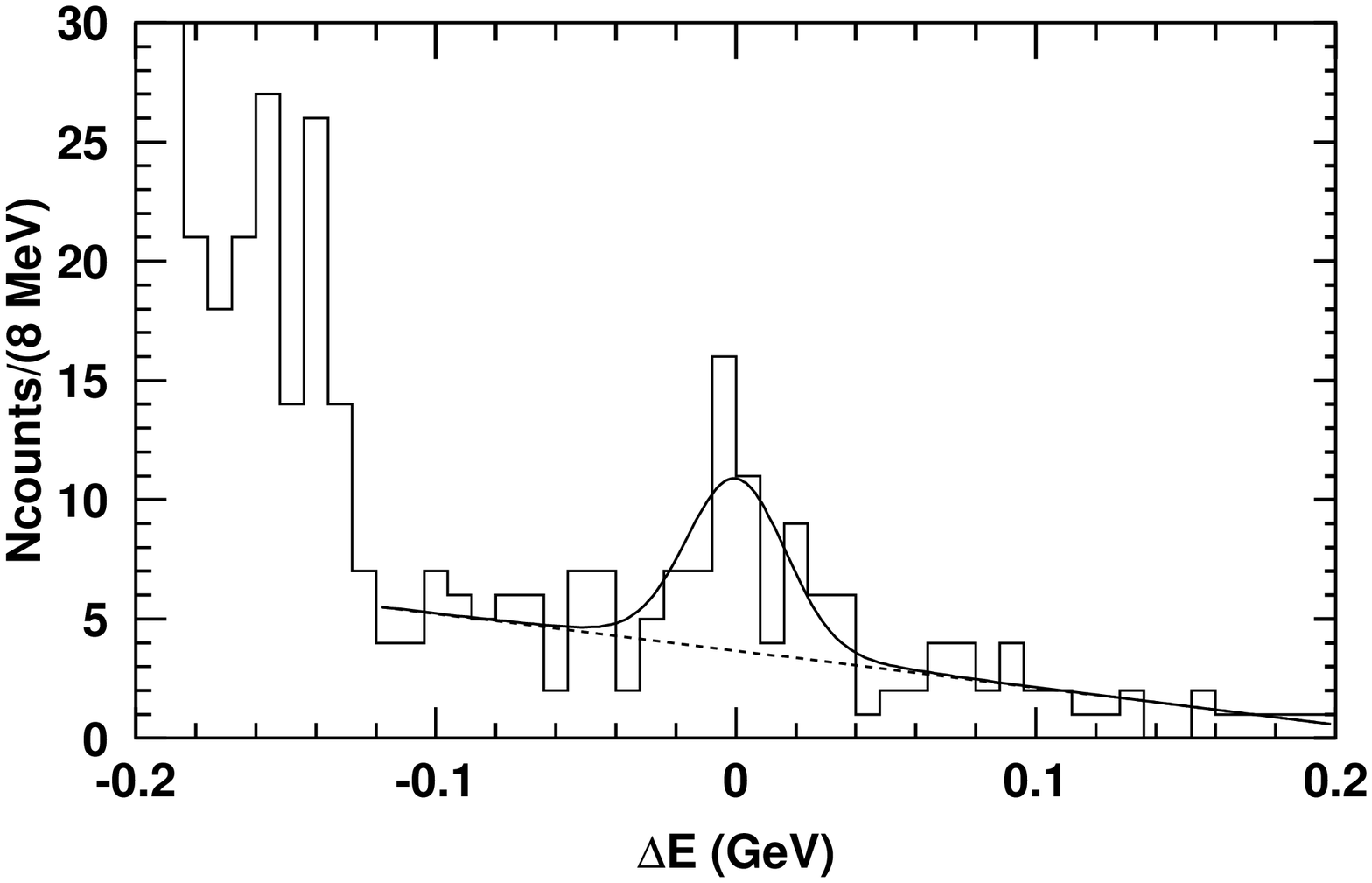} \hfill
    \includegraphics[height=3.0cm,width=8.0cm]{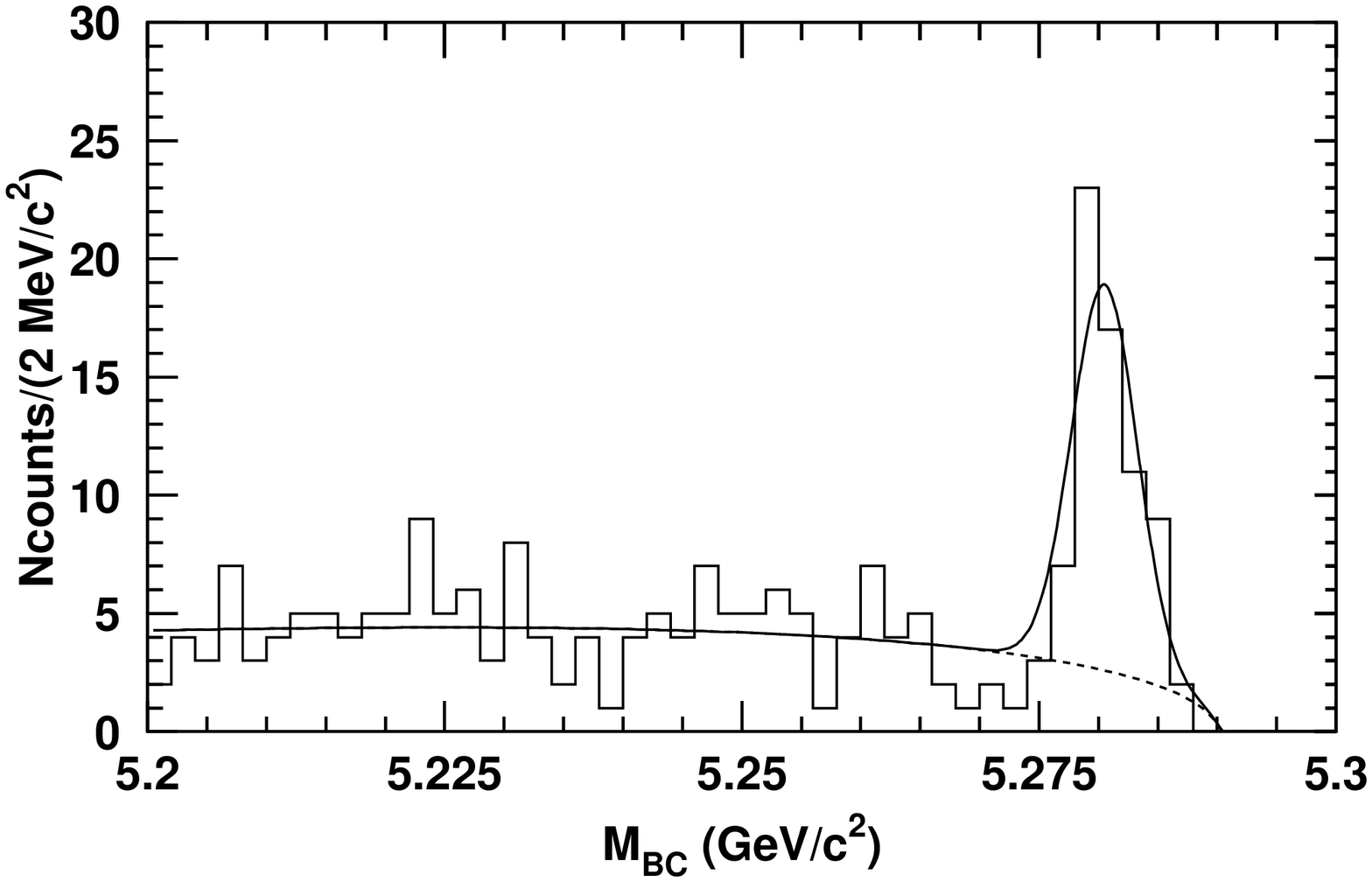} \\
    \includegraphics[height=3.0cm,width=8.0cm]{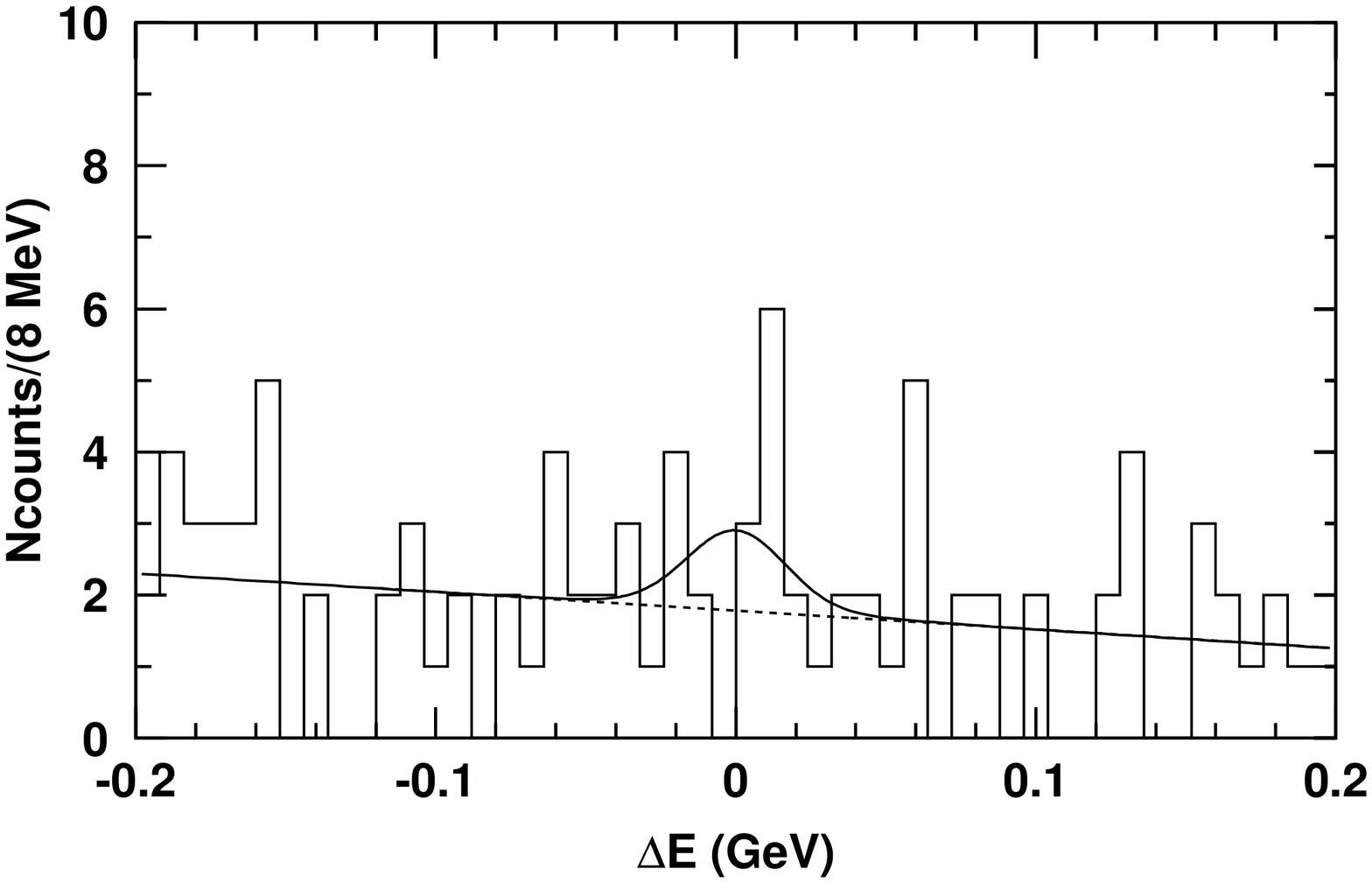} \hfill
    \includegraphics[height=3.0cm,width=8.0cm]{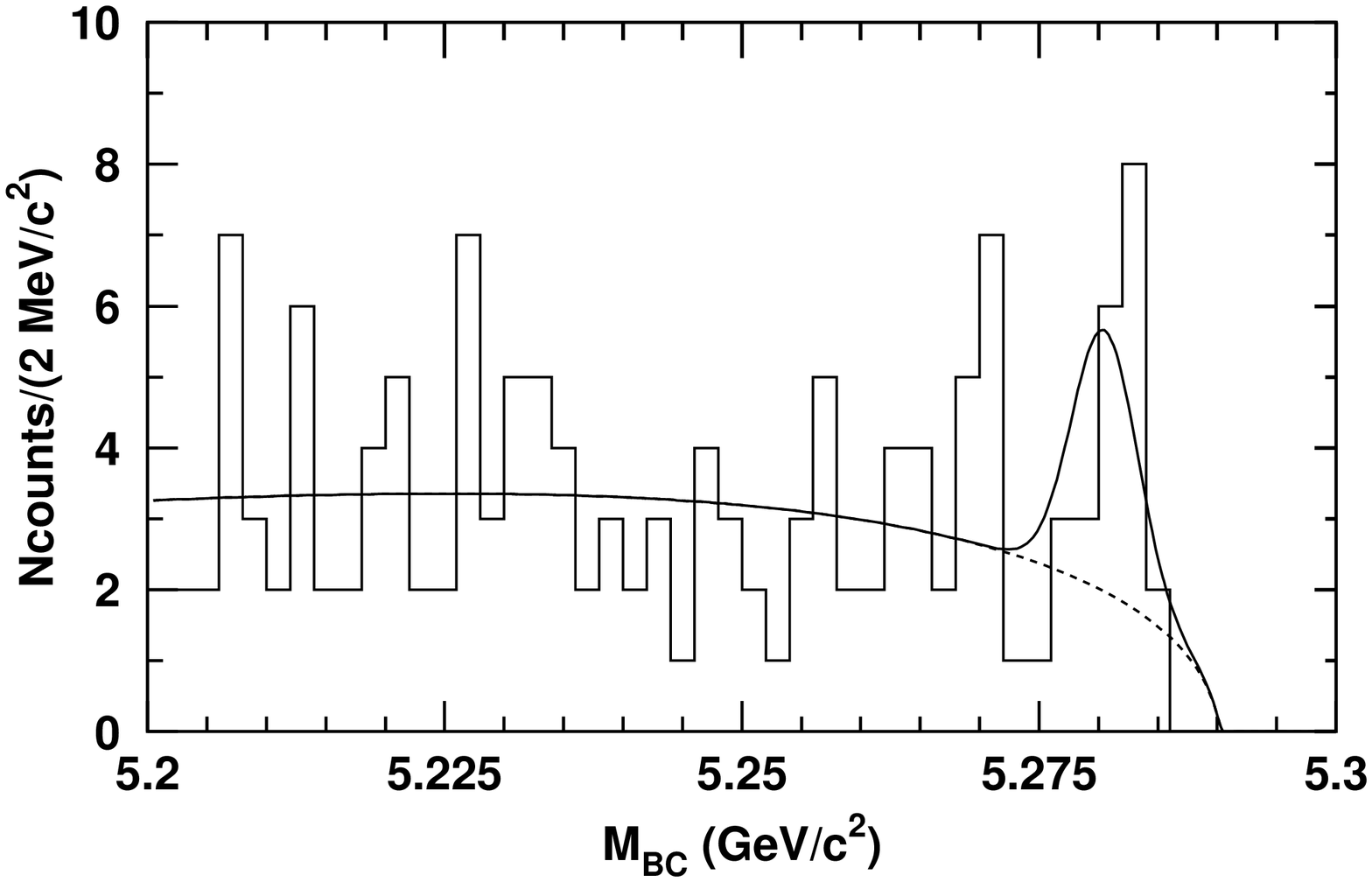} \\
    \includegraphics[height=3.0cm,width=8.0cm]{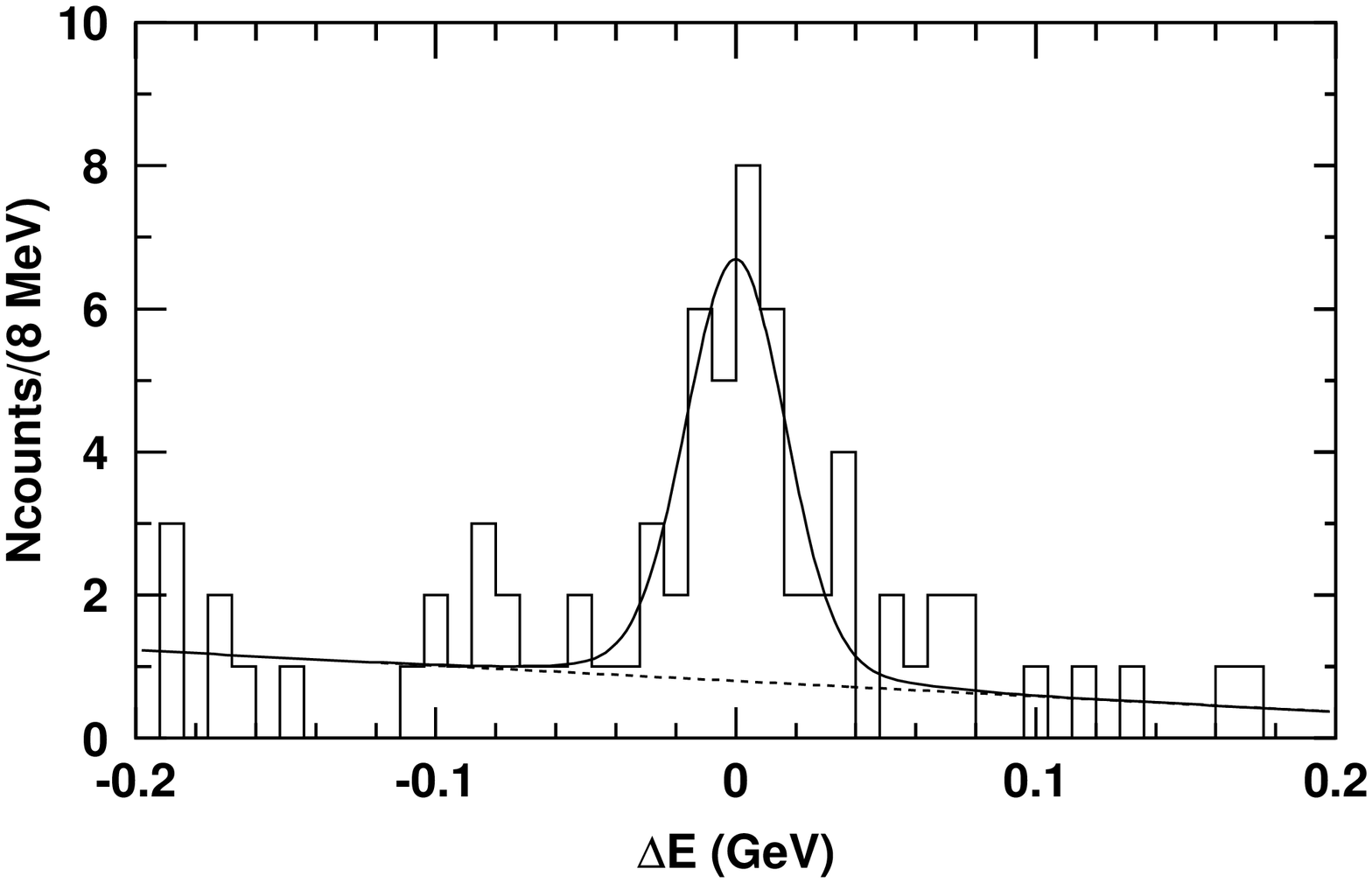} \hfill
    \includegraphics[height=3.0cm,width=8.0cm]{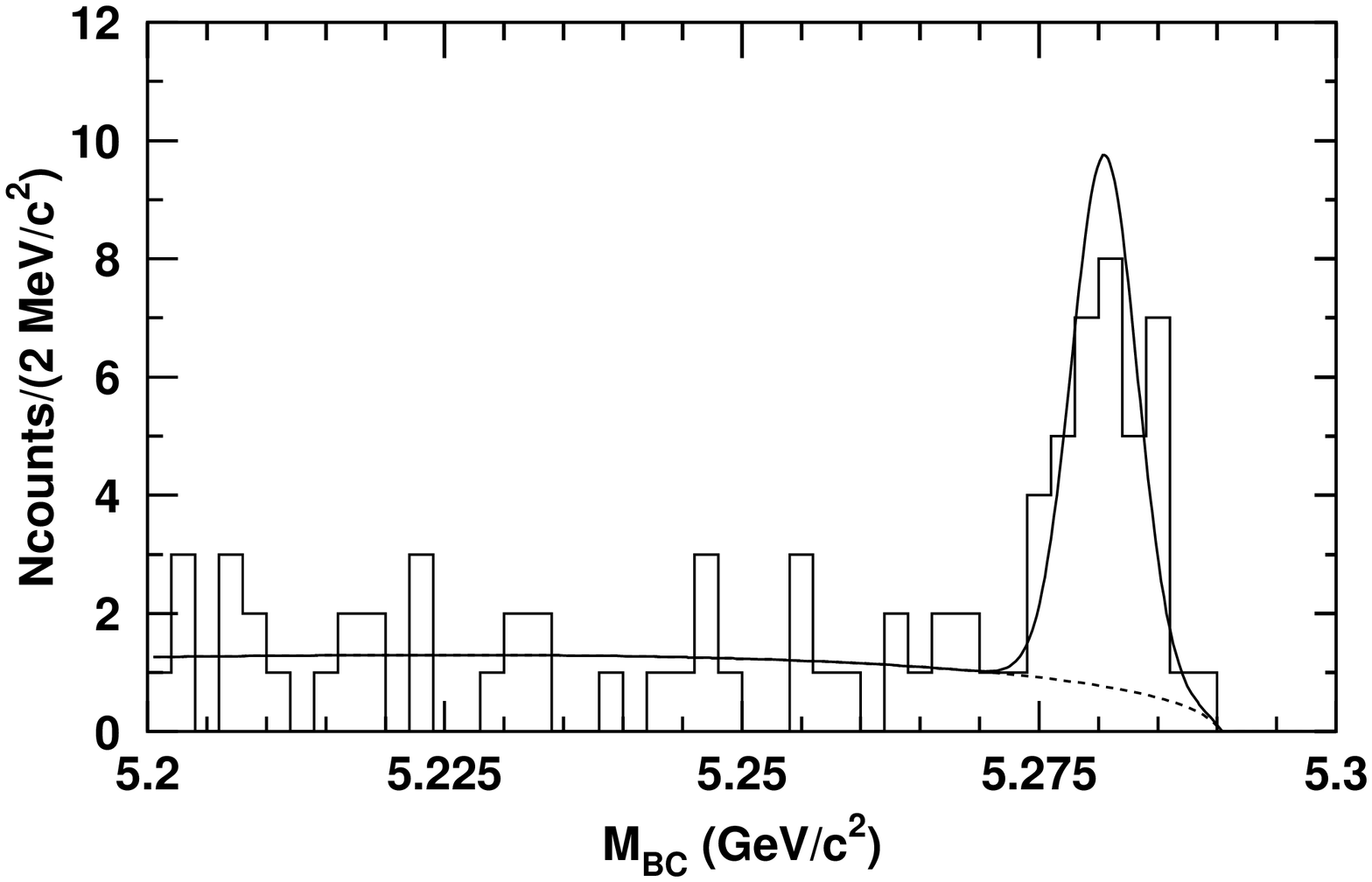} \\
    \includegraphics[height=3.0cm,width=8.0cm]{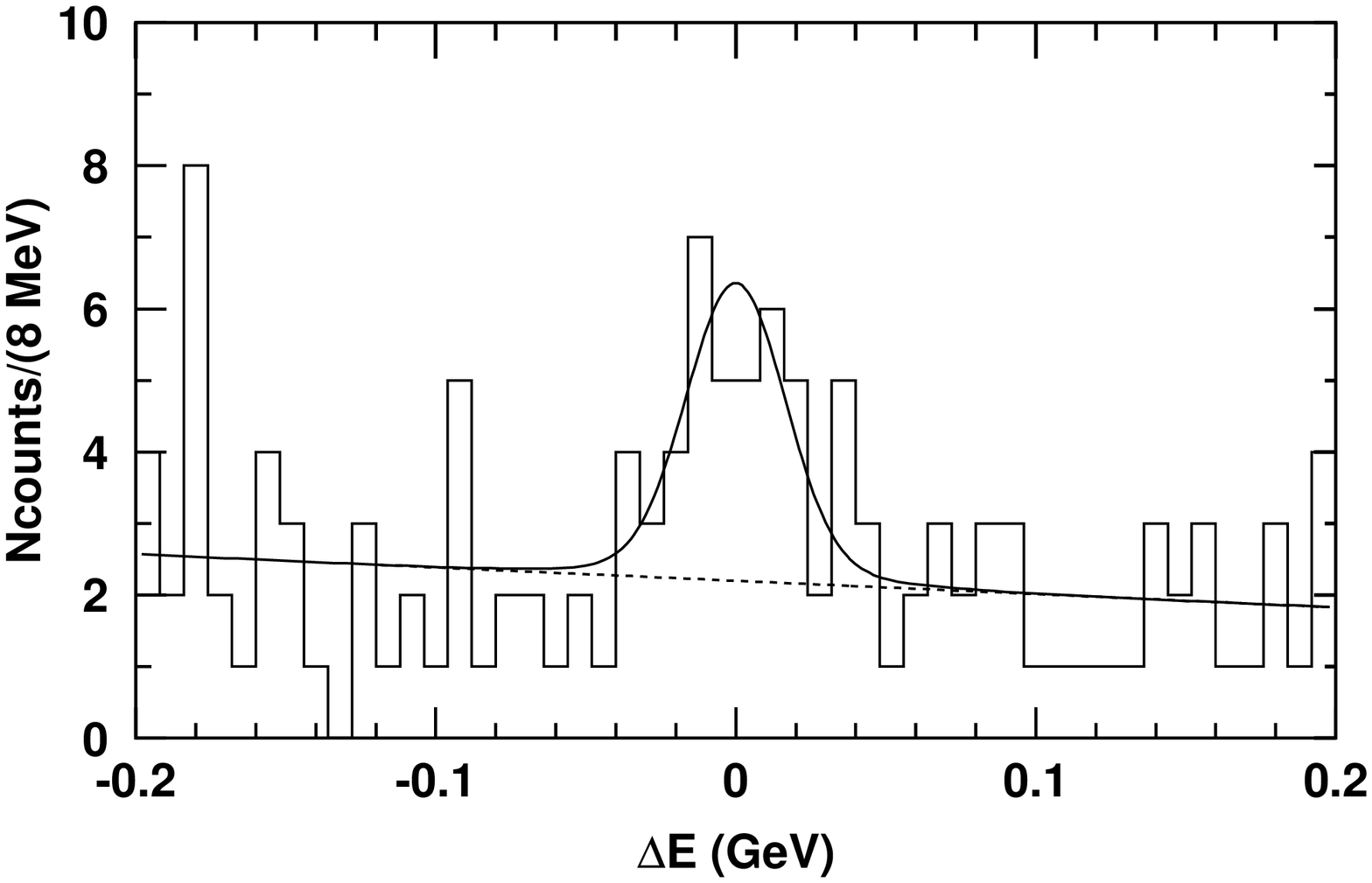} \hfill
    \includegraphics[height=3.0cm,width=8.0cm]{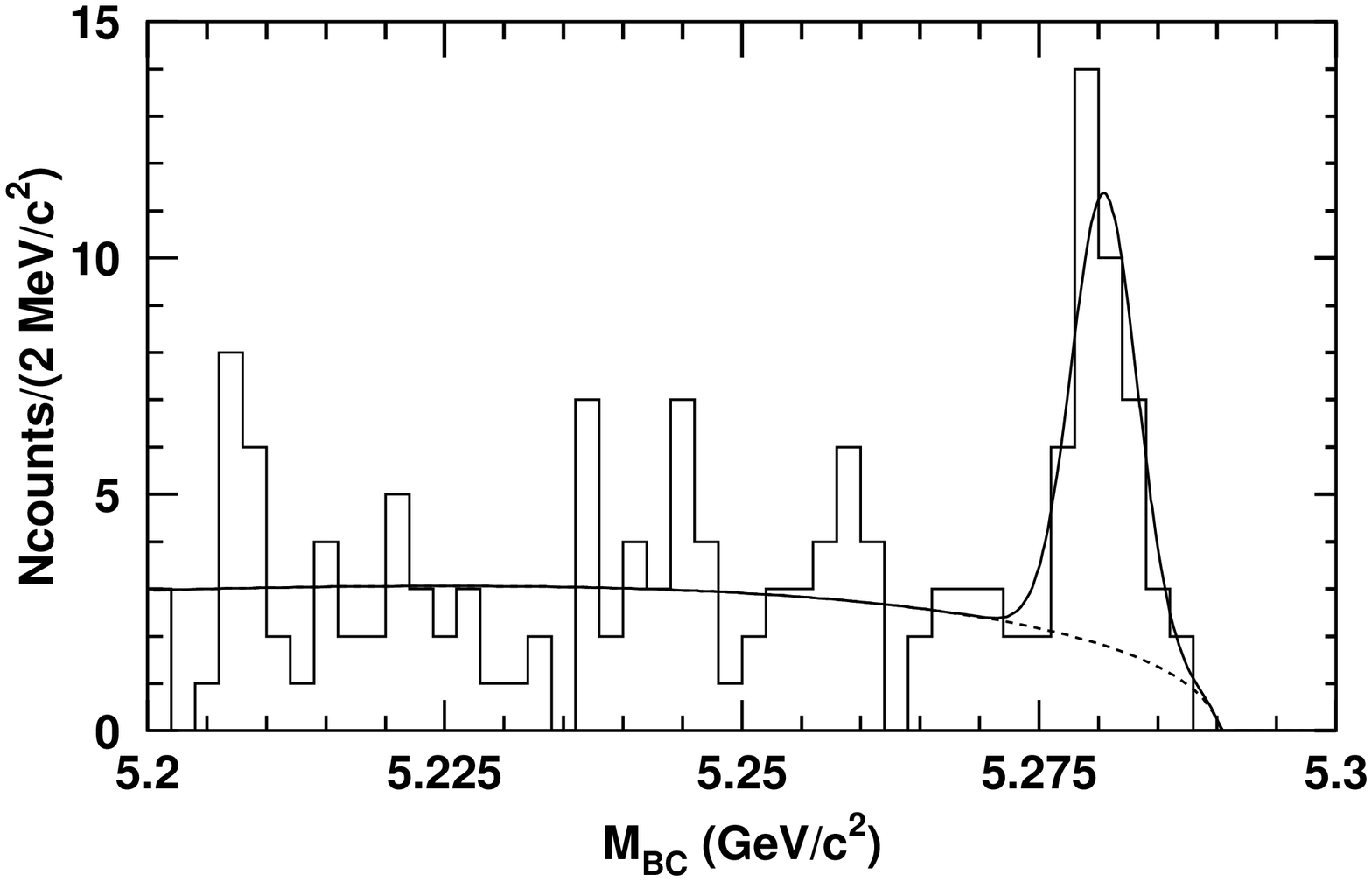} \\
    \includegraphics[height=3.0cm,width=8.0cm]{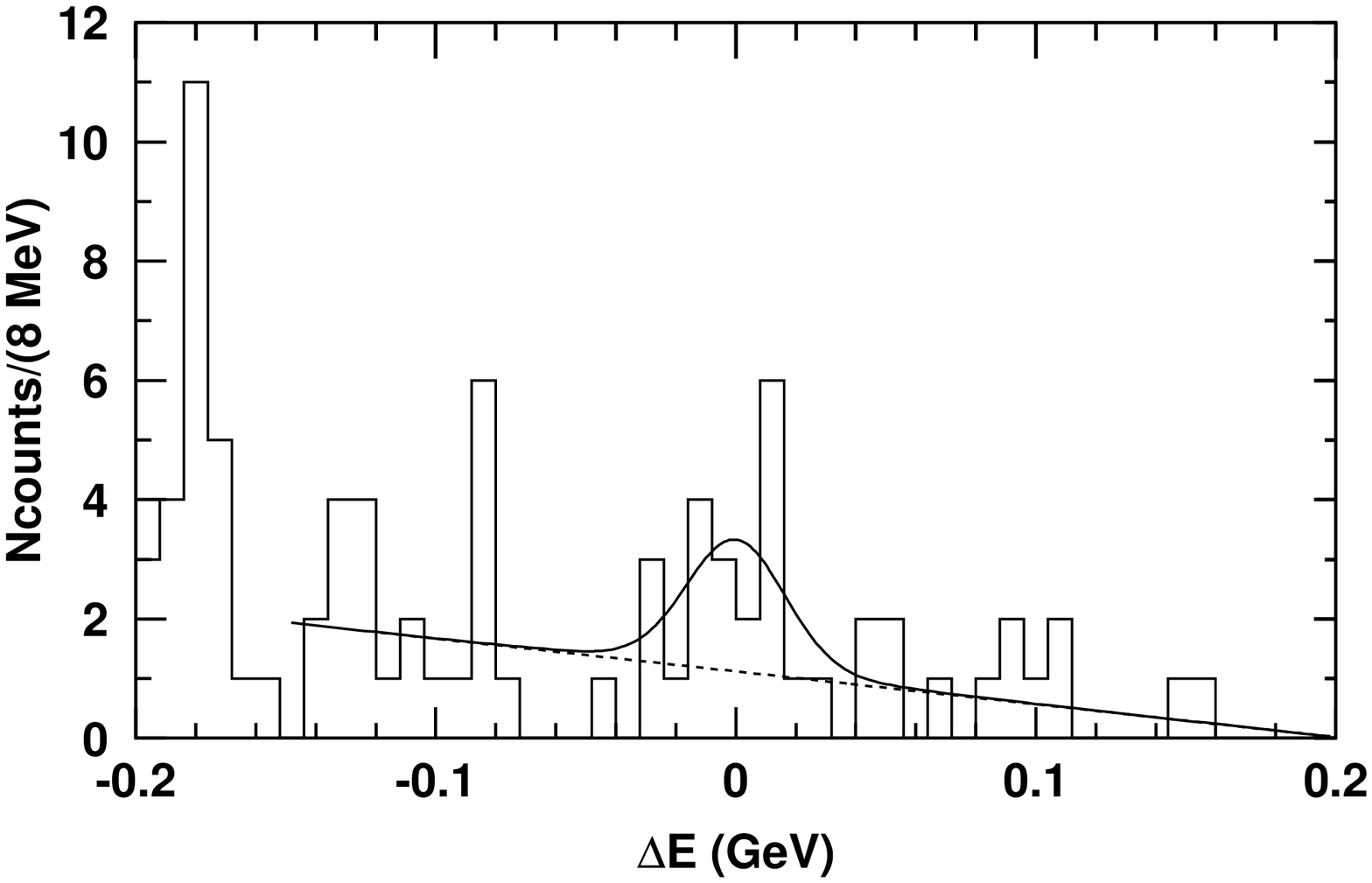} \hfill
    \includegraphics[height=3.0cm,width=8.0cm]{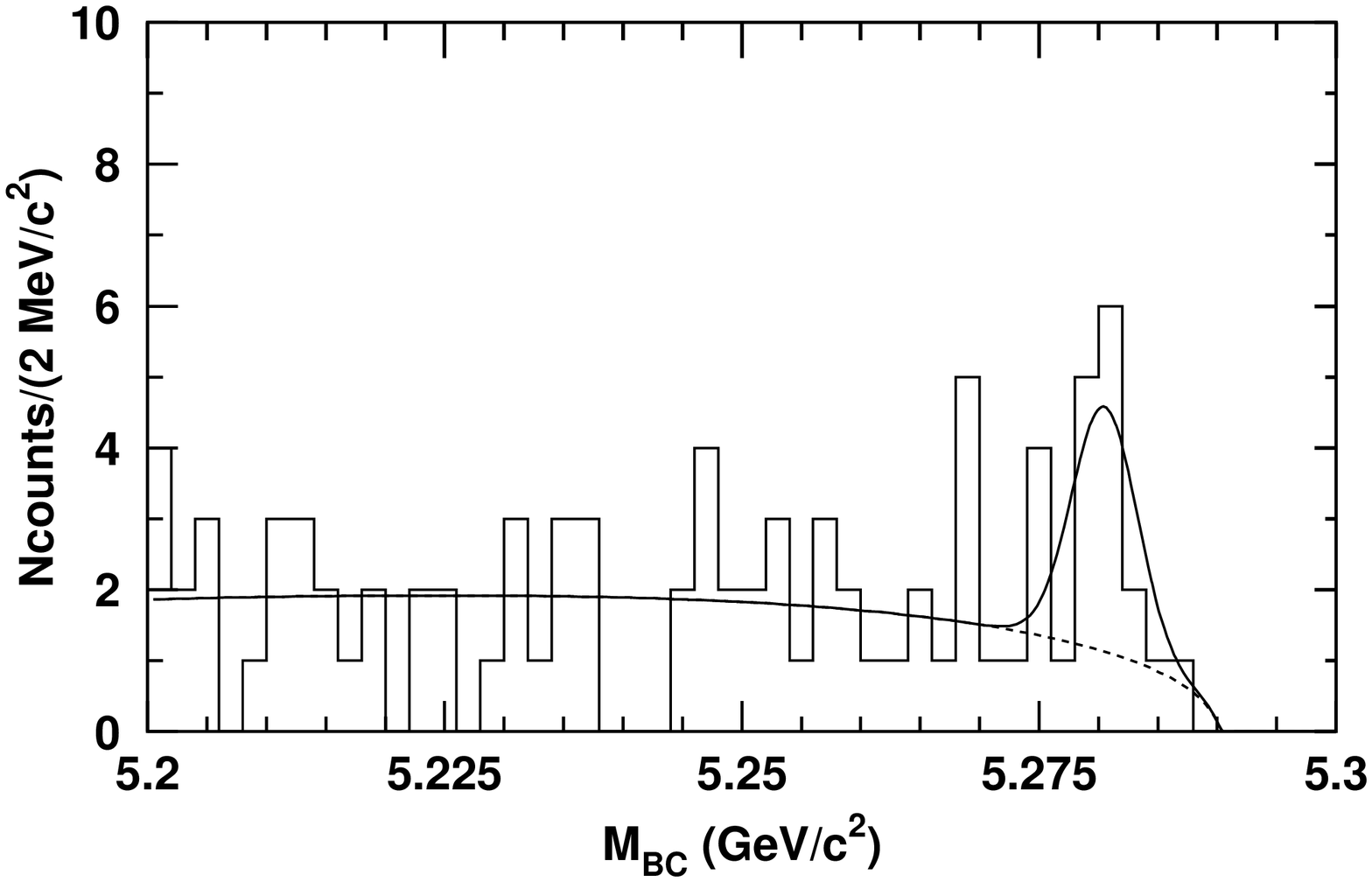} \\
    \includegraphics[height=3.0cm,width=8.0cm]{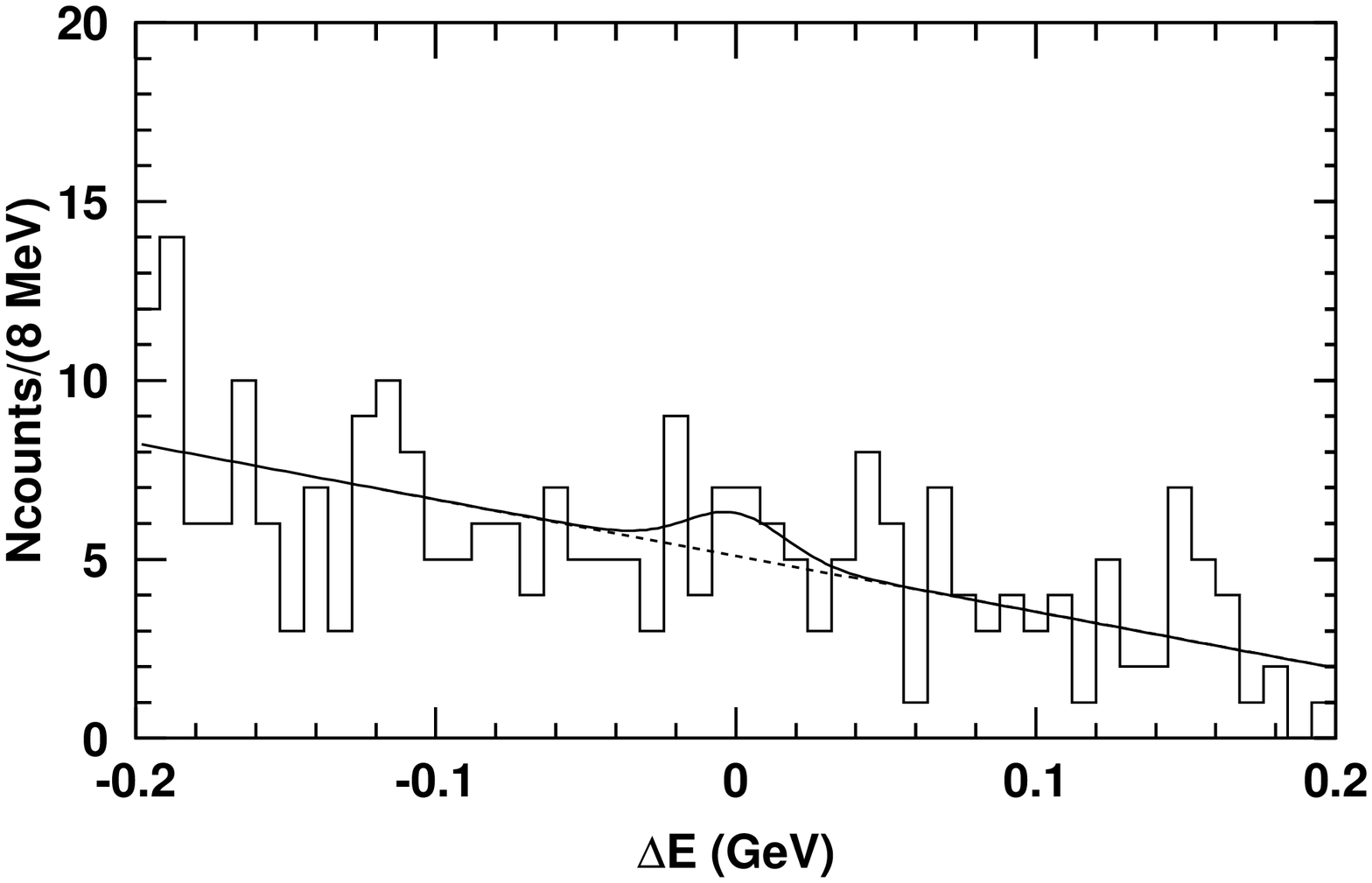} \hfill
    \includegraphics[height=3.0cm,width=8.0cm]{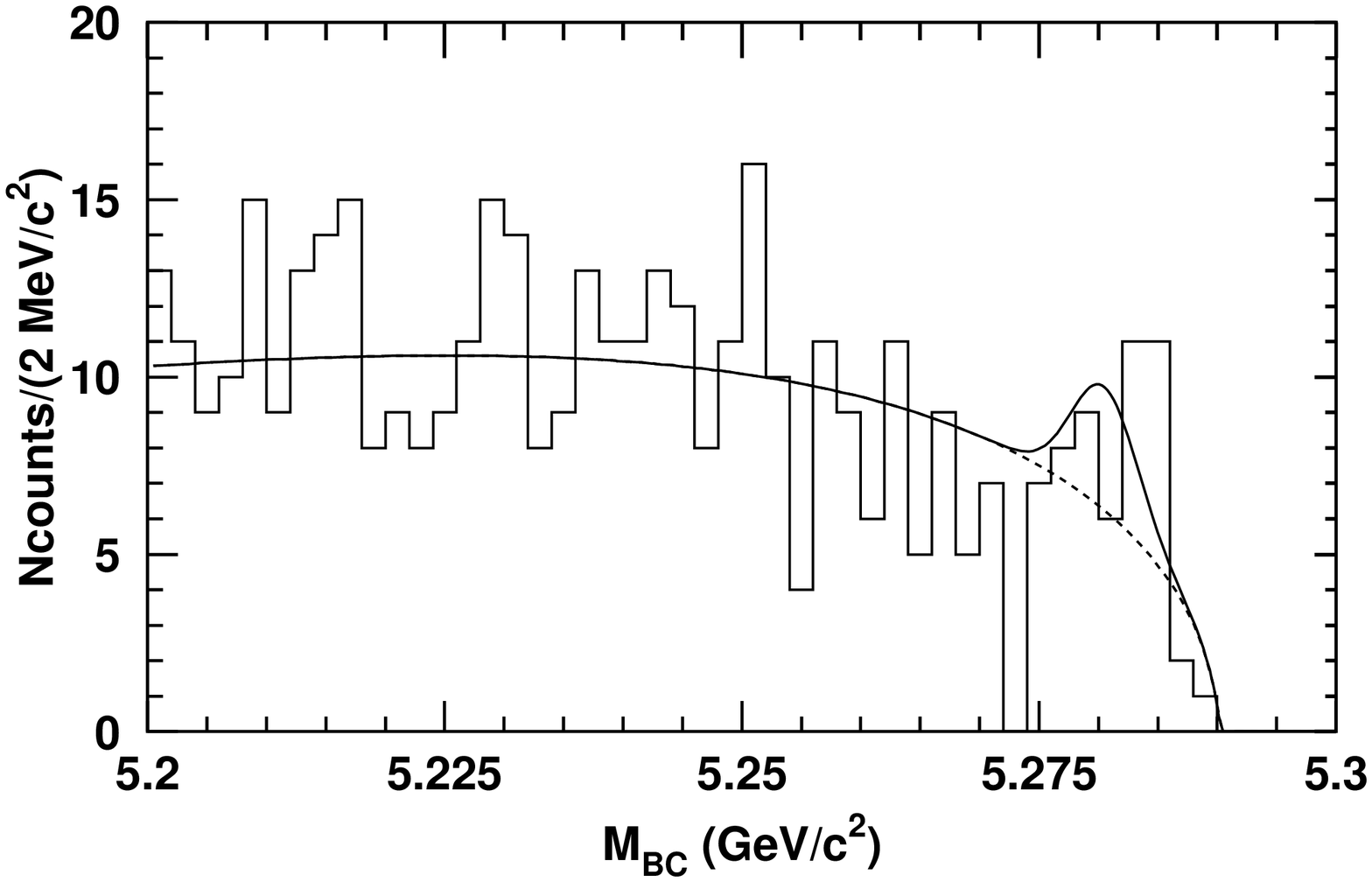}

  \caption{The $\Delta E$ (left) and $M_{BC}$ (right) distributions for different 
           regions of the $K^+\pi^+\pi^-$ Dalitz plot. 
           The plots from top to bottom correspond to regions I to VII, respectively.}
  \label{pp_reso}
\end{figure}

%%%%%%%%%%%%%%%%%%%%%%%%%%%%%%%%    Plots for KKK section    %%%%%%%%%%%%%%%%%%%%%%%%%%

\begin{figure}[htb]
  \centering
  \includegraphics[height=10.0cm,width=12.0cm]{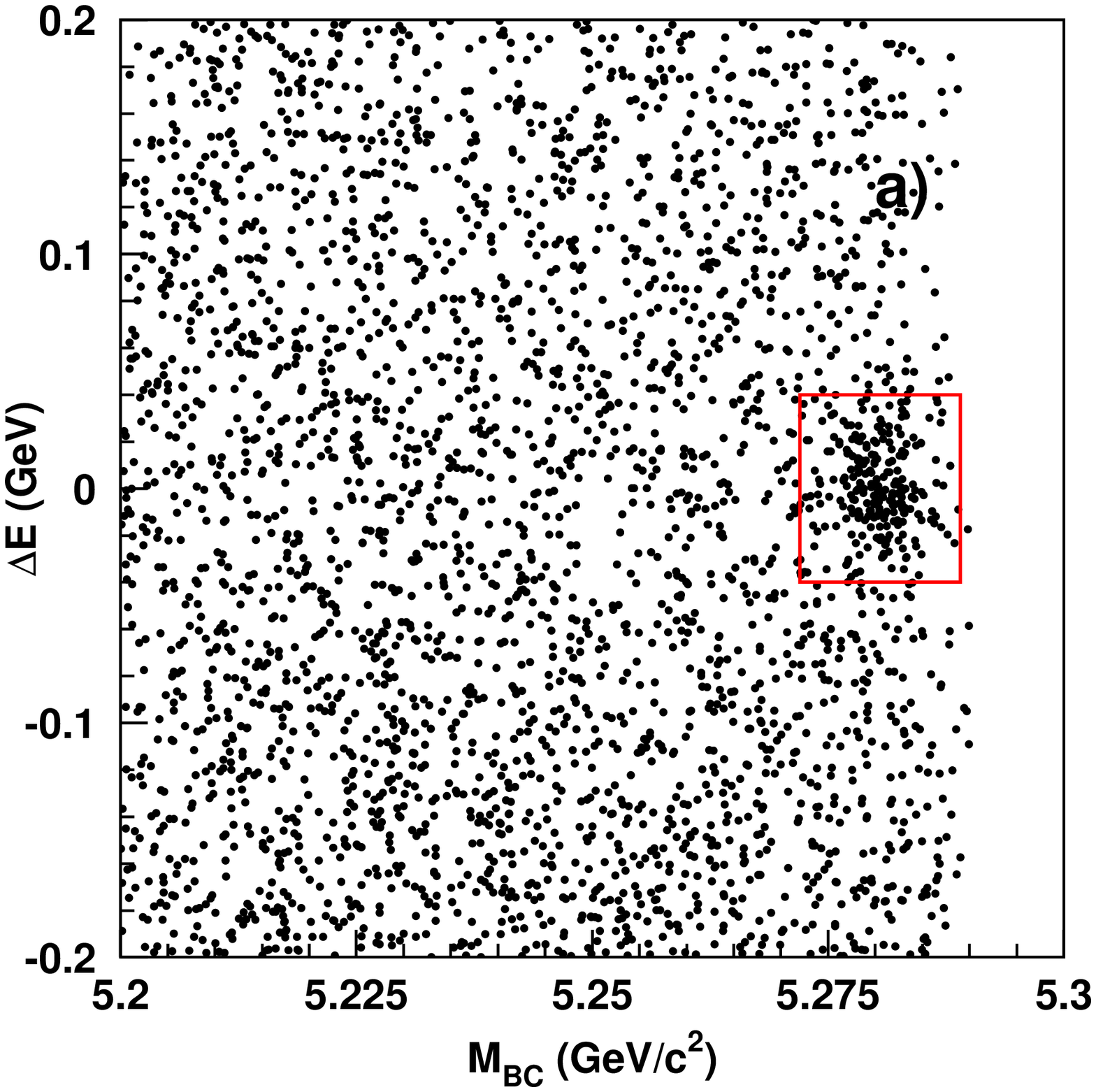} \hfill
  \includegraphics[height=10.0cm,width=12.0cm]{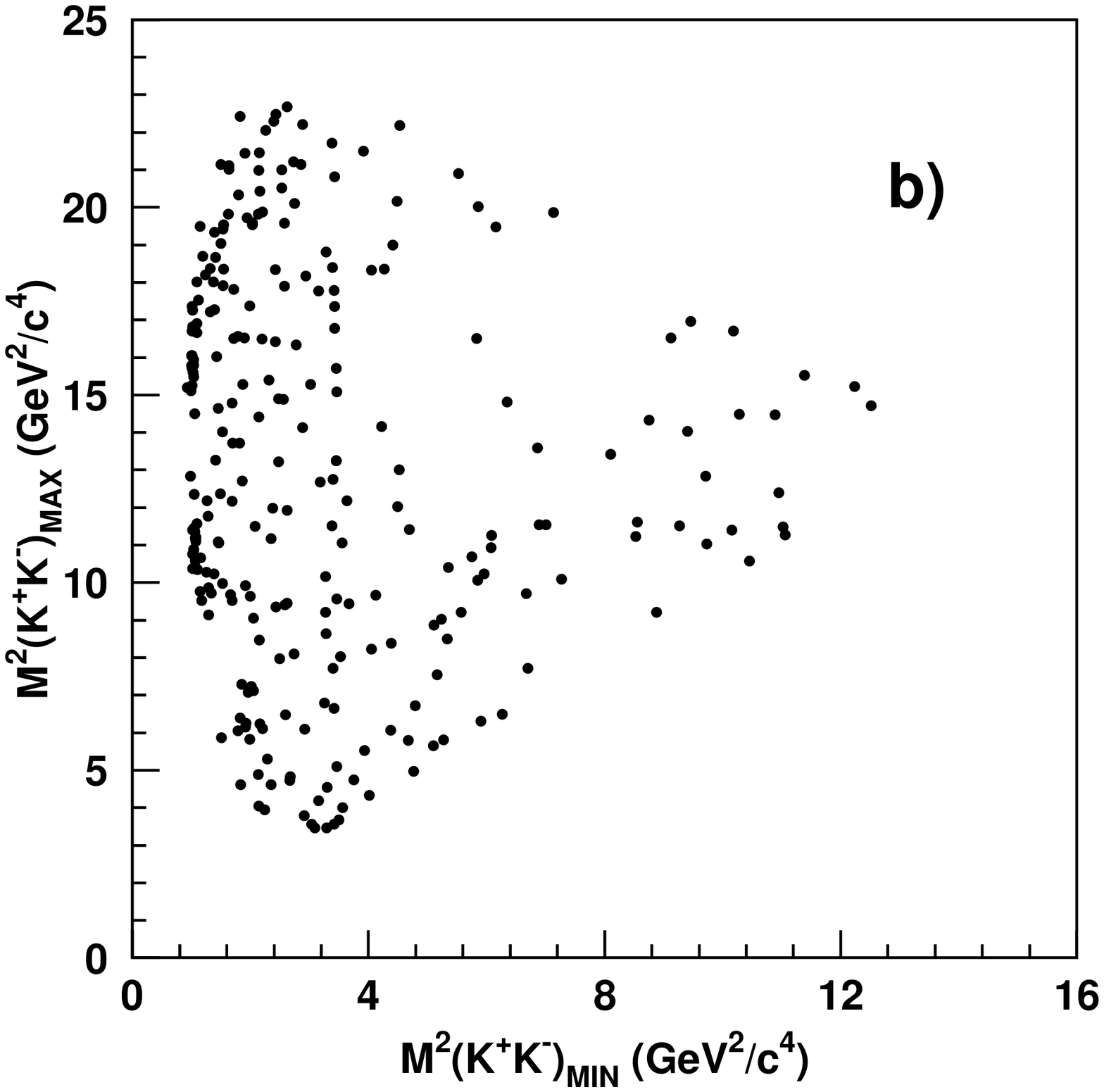}

  \centering
  \caption{ {\bf a)} The $\Delta E$ versus $M_{BC}$ plot for all 
                     $B^+ \to K^+K^+K^-$ candidates.
            {\bf b)} The Dalitz plot for $B^+ \to K^+K^-K^+$ candidates
                     from the $B$ signal region inside the box in a).}
  \label{kkk_total}
\end{figure}

\begin{figure}[htb]
  \hspace*{-0.1cm}\includegraphics[height=8.0cm,width=8.5cm]{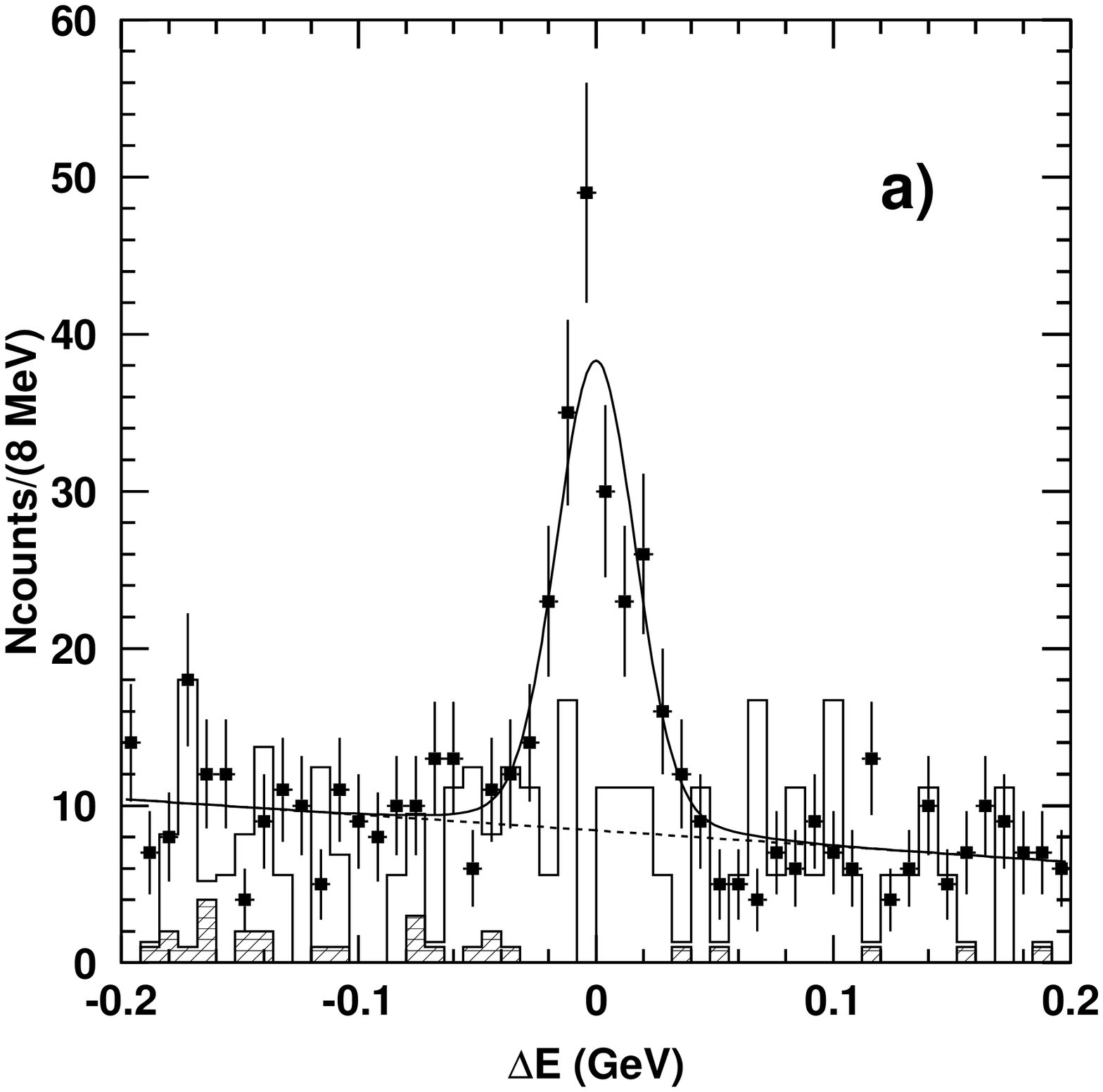} \hfill
  \hspace*{-0.7cm}\includegraphics[height=8.0cm,width=8.5cm]{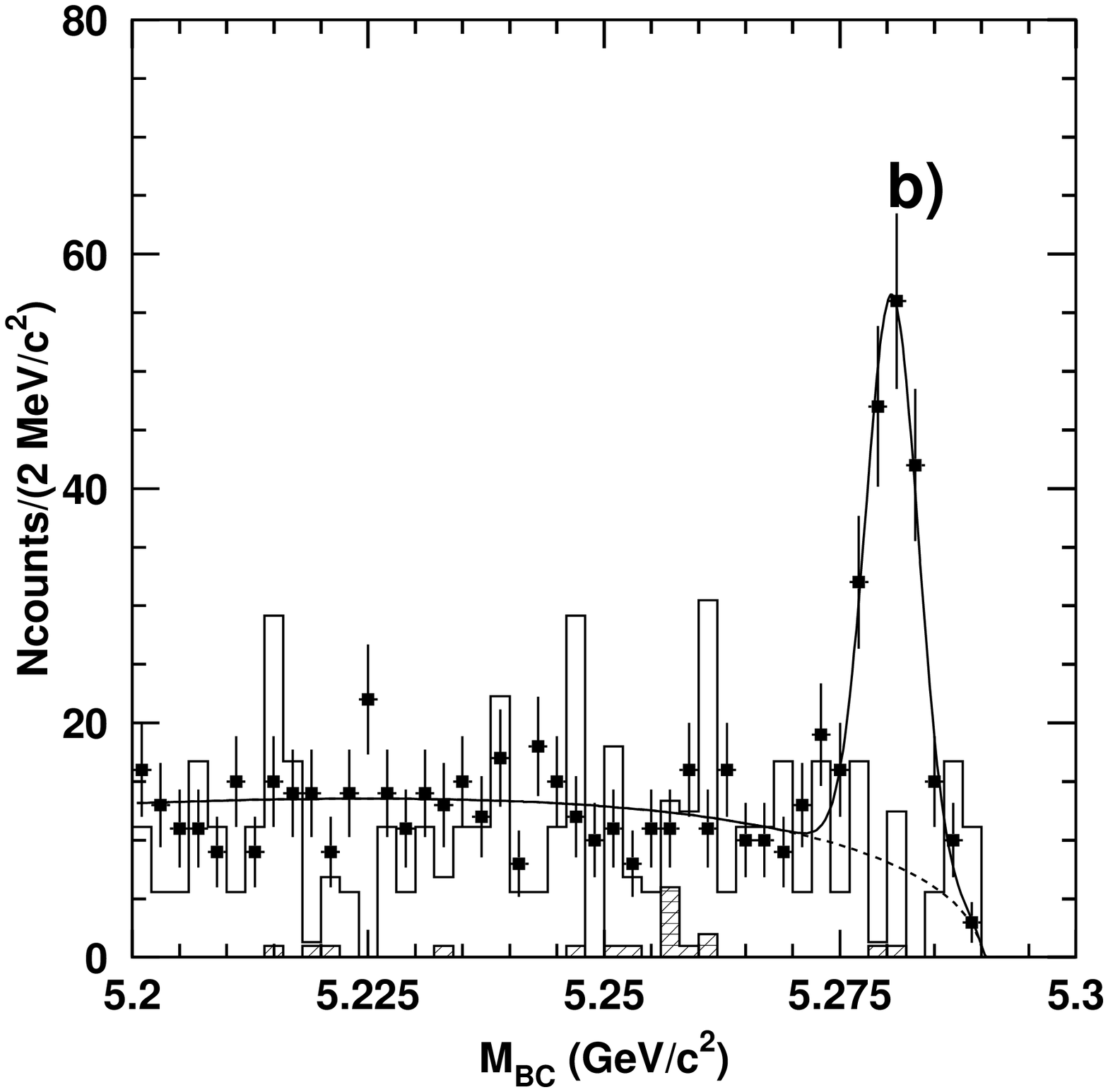}

  \caption{The $\Delta E$ - a) and $M_{BC}$ - b) distributions for the
           $B^+ \to K^+K^+K^-$ final state.
           Candidates consistent with $B^+ \to D^o_{CP}\pi^+$ or 
           $B^+ \to D^o_{CP} K^+$ where $D^o_{CP}\to K^+K^-$ are excluded. 
           Points are data, open histograms are the proper sum of the 
           off-resonance data and $B\bar{B}$ 
           Monte Carlo, and hatched histograms show the contribution of 
           $B\bar{B}$ Monte Carlo only. The curves show the fit to the data.}
  \label{kkk_mbde}
\end{figure}

\begin{figure}[htb]
  \hspace*{-0.1cm}\includegraphics[height=8.0cm,width=8.5cm]{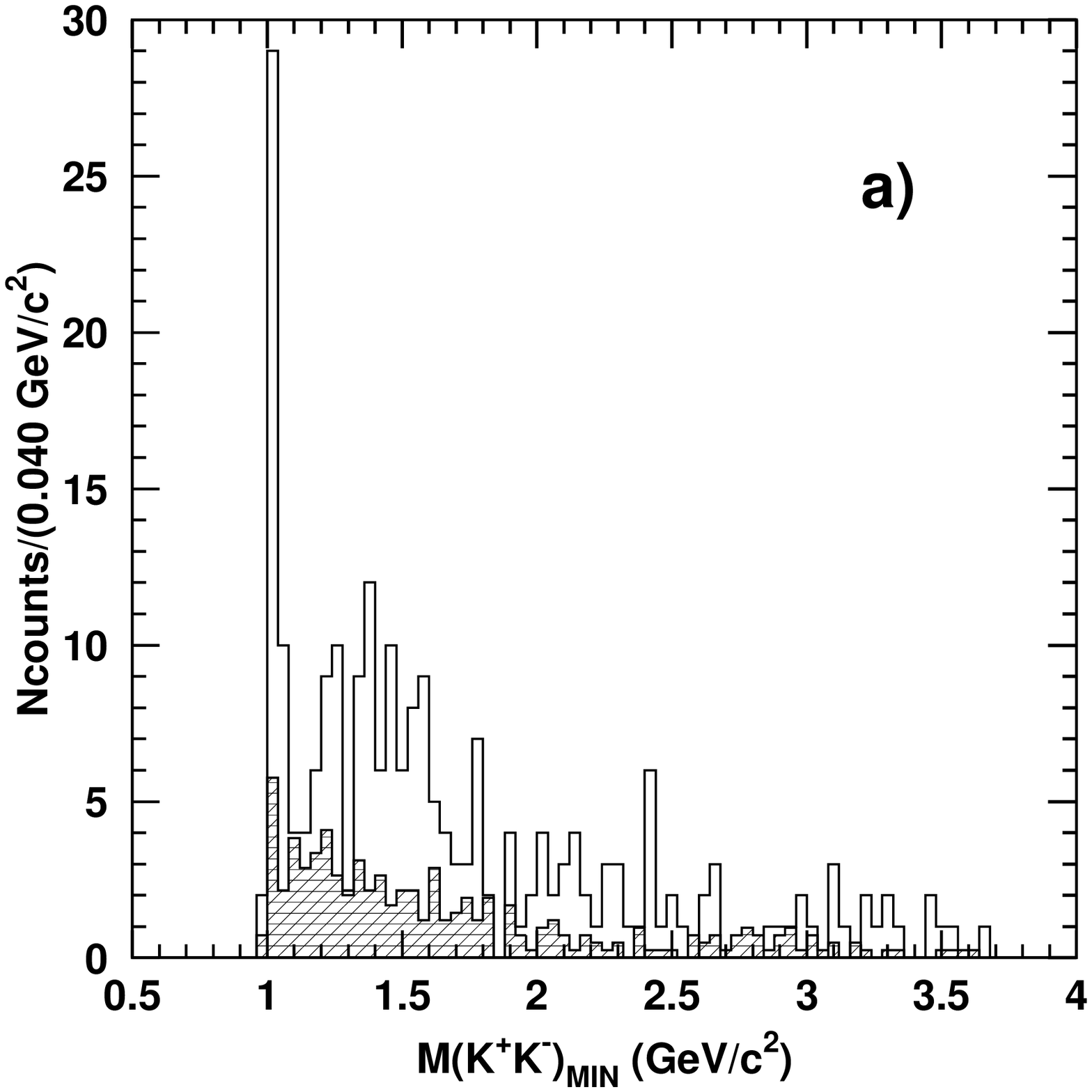} \hfill
  \hspace*{-0.7cm}\includegraphics[height=8.0cm,width=8.5cm]{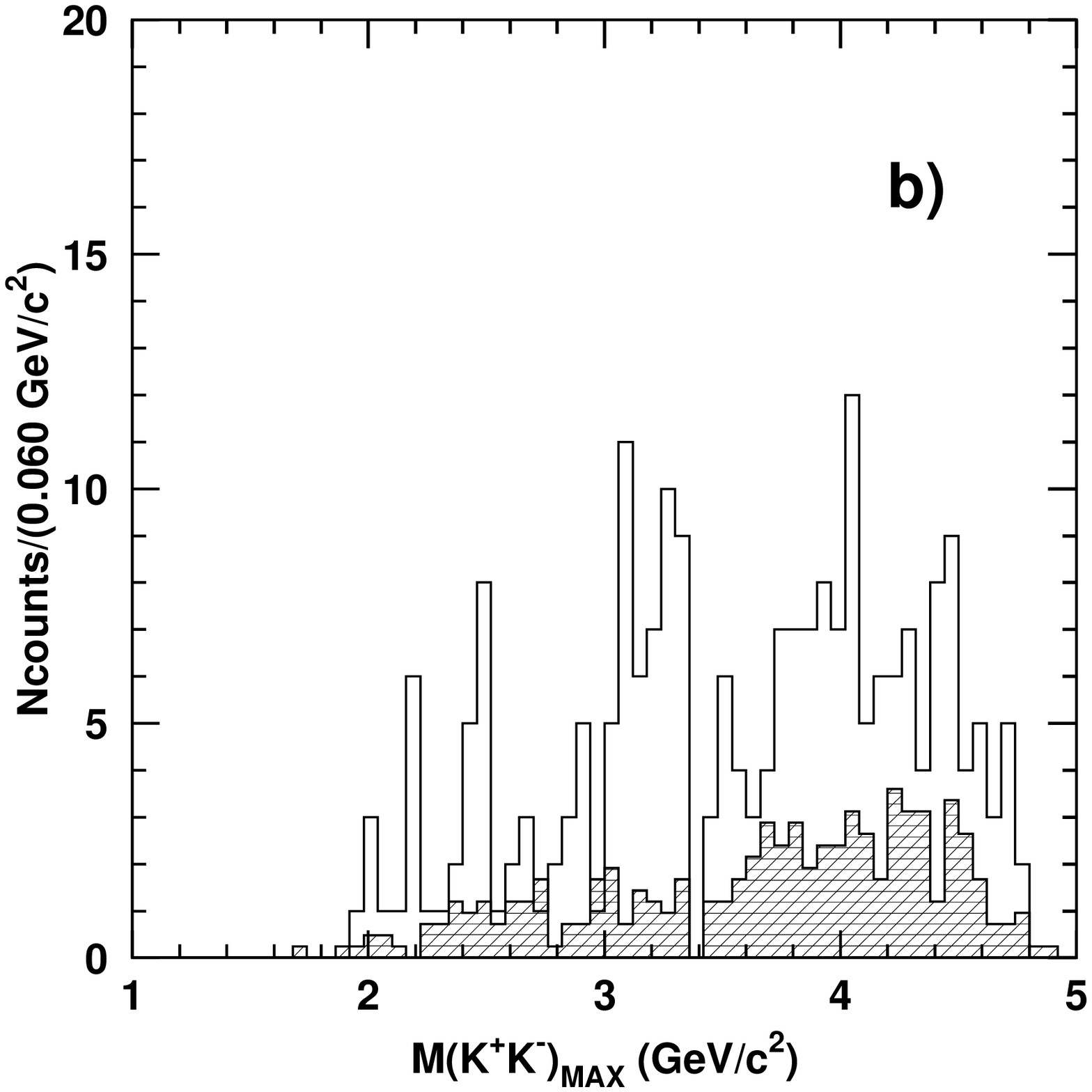}

  \caption{ The $K^+K^-$ invariant mass spectrum for $B^+\to K^+K^-K^+$ candidates. 
            Open histogram for candidates from the $B$ signal region, and
            hatched histogram for candidates from the $\Delta E$ sidebands.
           {\bf a)} The $K^+K^-$ combination with the smaller invariant mass.
           {\bf b)} The $K^+K^-$ combination with the larger  invariant mass.}
  \label{kkmass}
\end{figure}

\begin{figure}[hbt]
    \includegraphics[height=4.0cm,width=8.0cm]{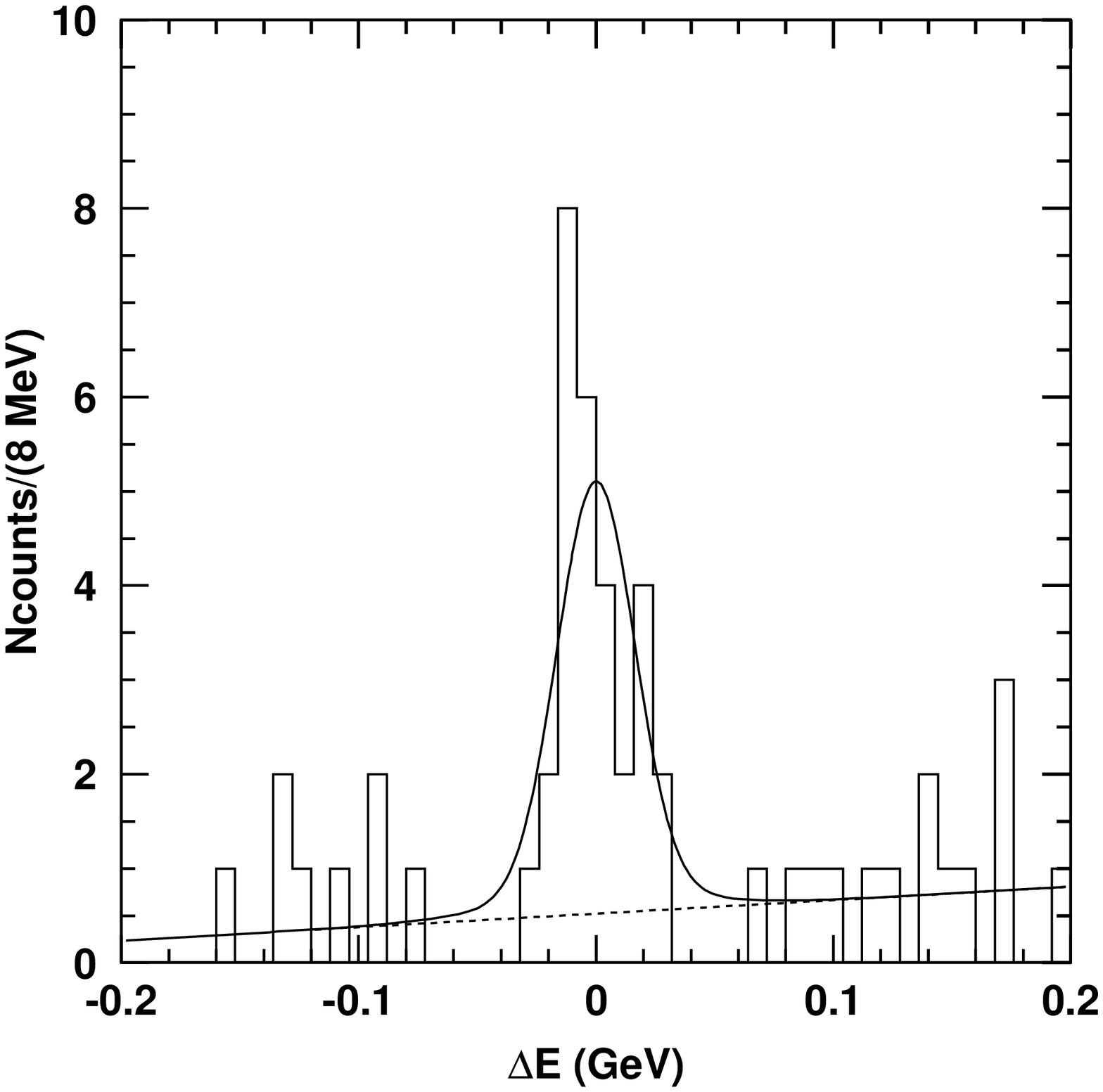} \hfill
    \includegraphics[height=4.0cm,width=8.0cm]{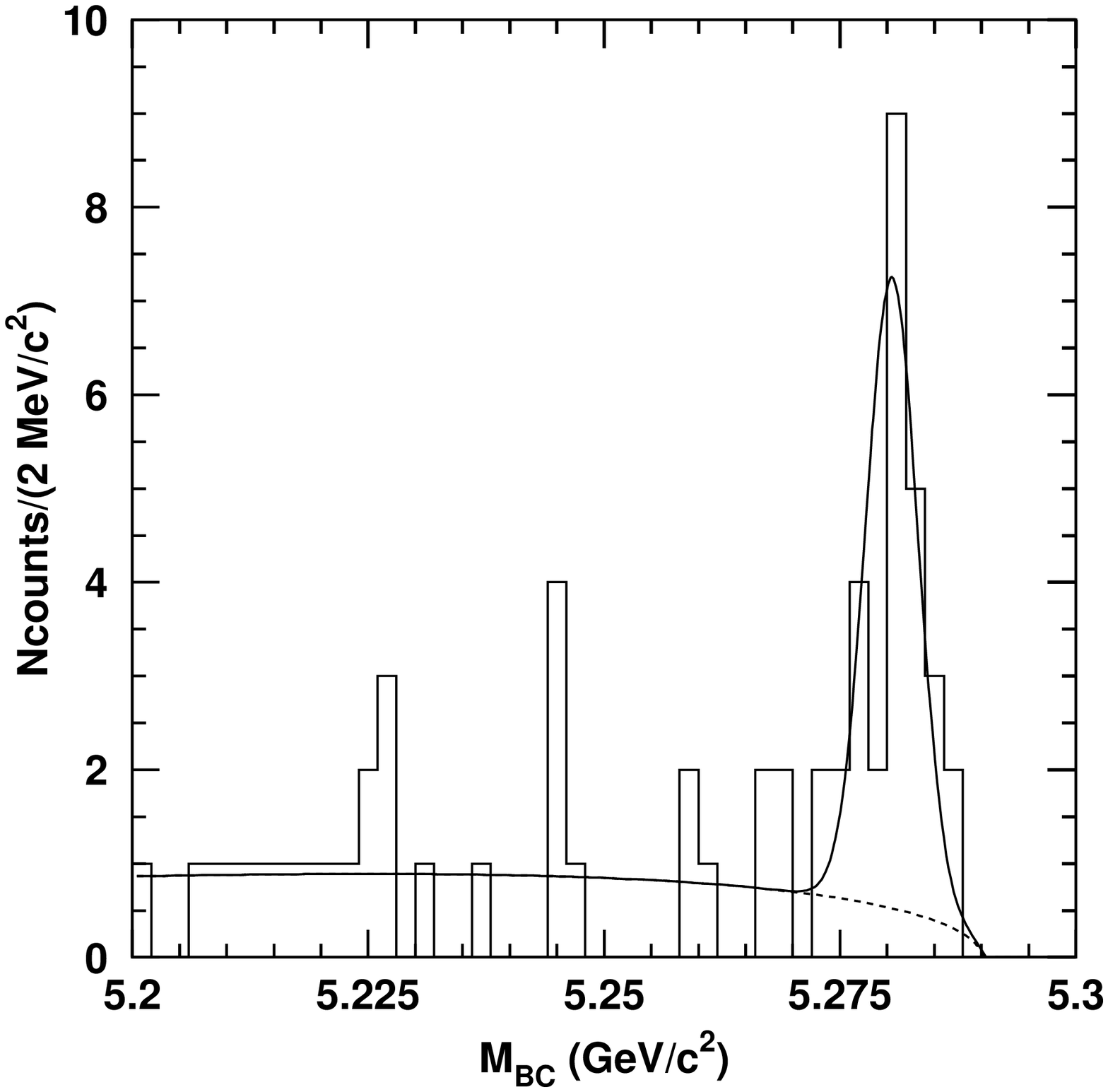} \\
    \includegraphics[height=4.0cm,width=8.0cm]{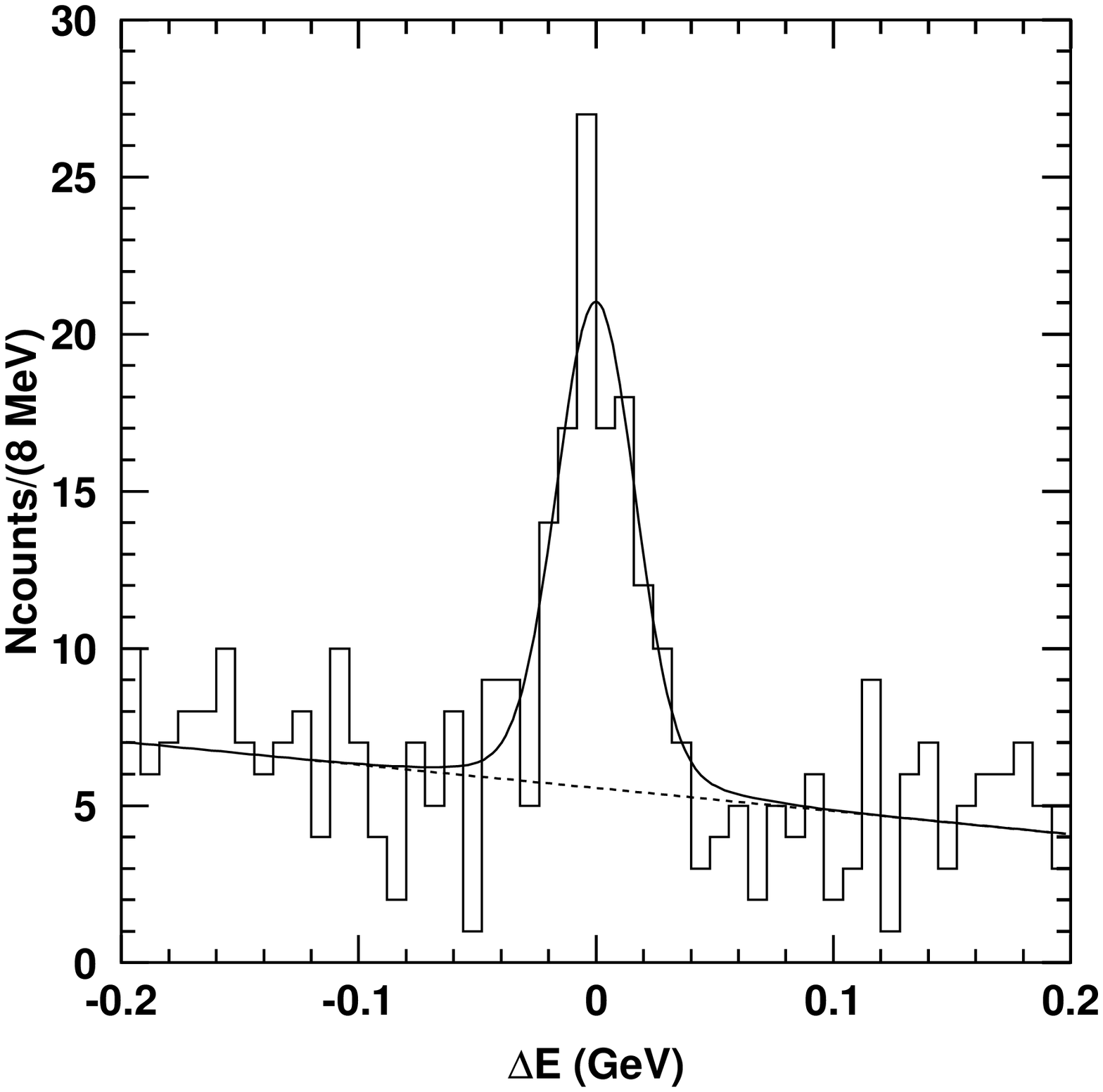} \hfill
    \includegraphics[height=4.0cm,width=8.0cm]{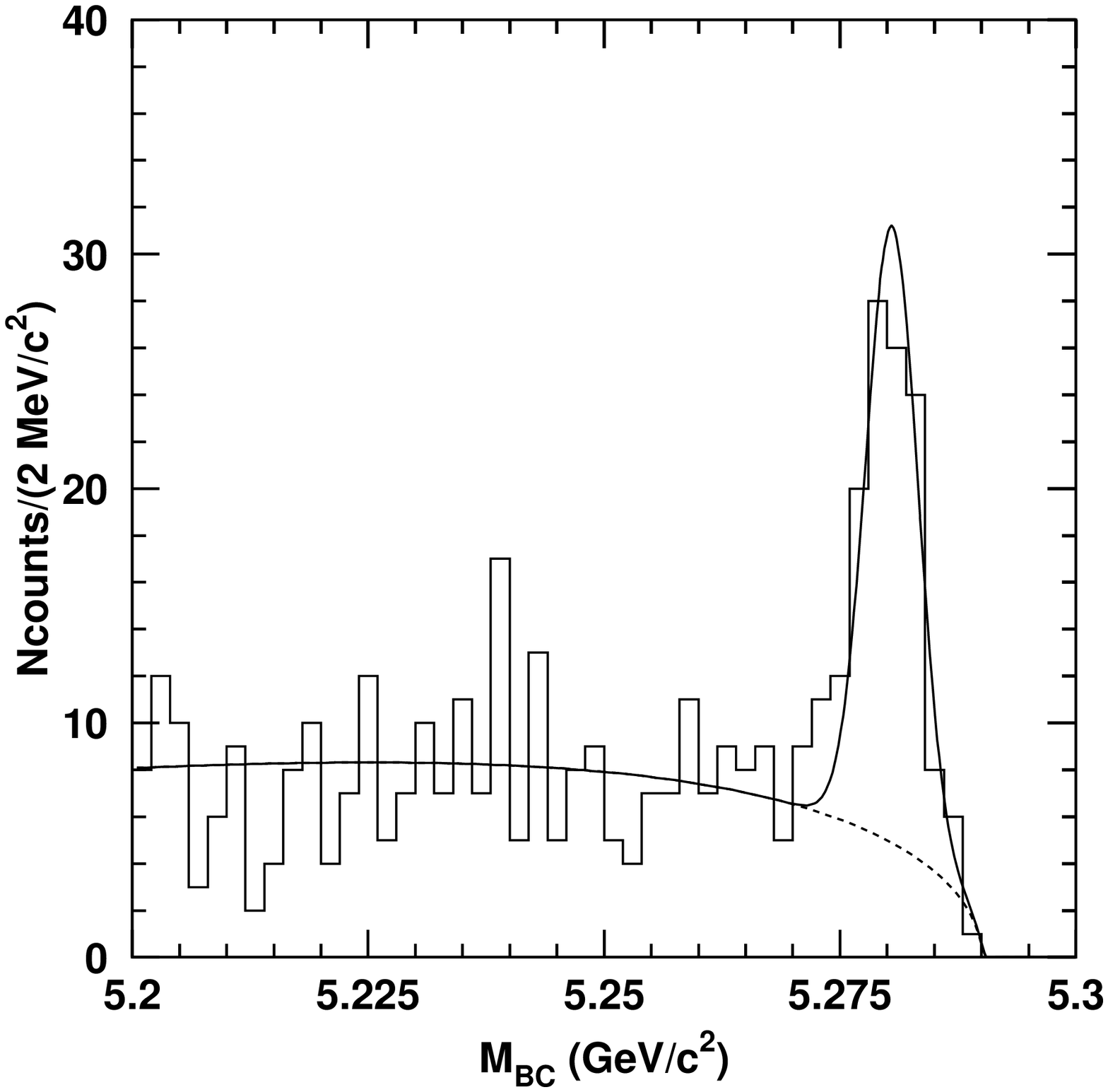} \\
    \includegraphics[height=4.0cm,width=8.0cm]{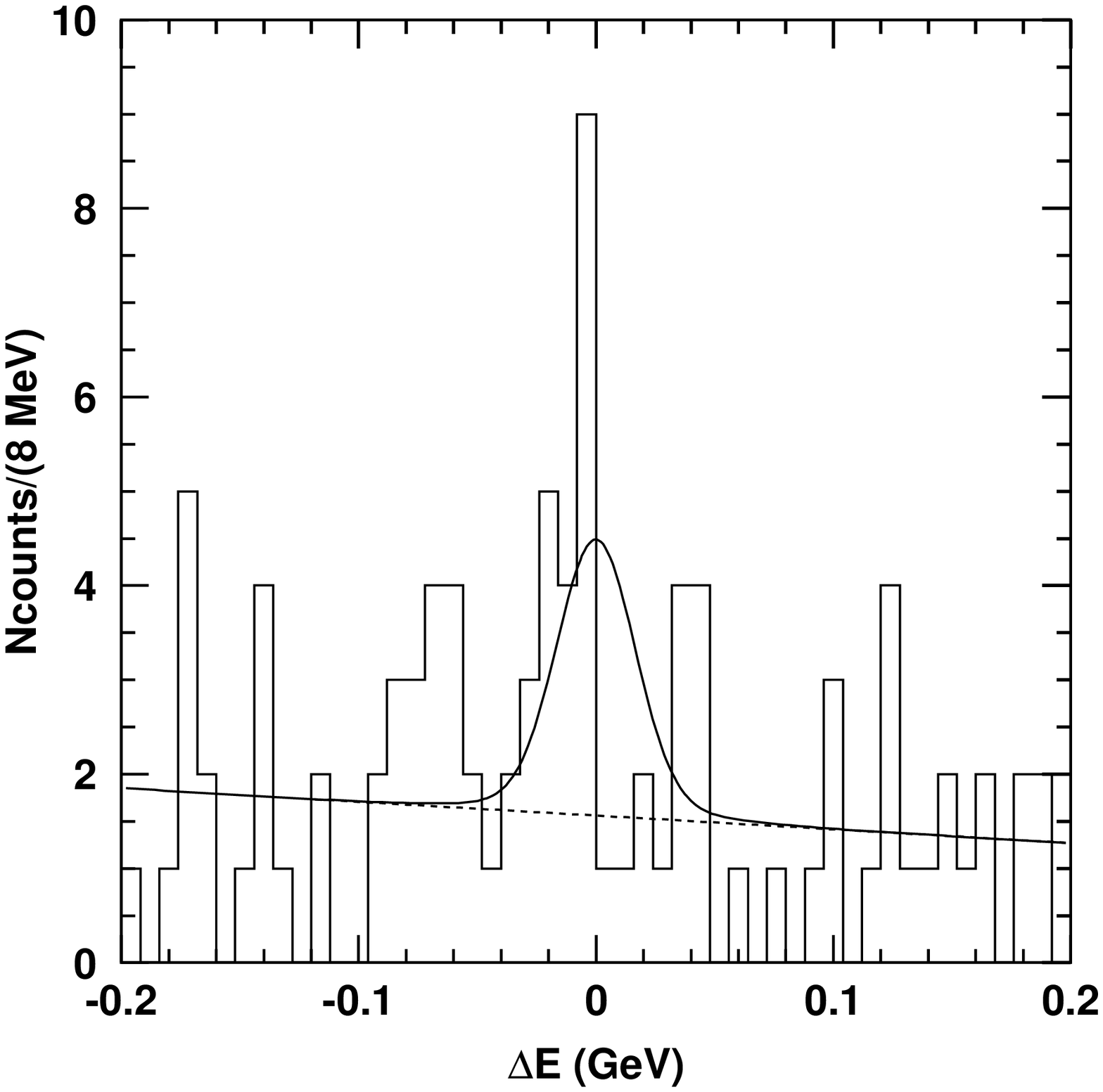} \hfill
    \includegraphics[height=4.0cm,width=8.0cm]{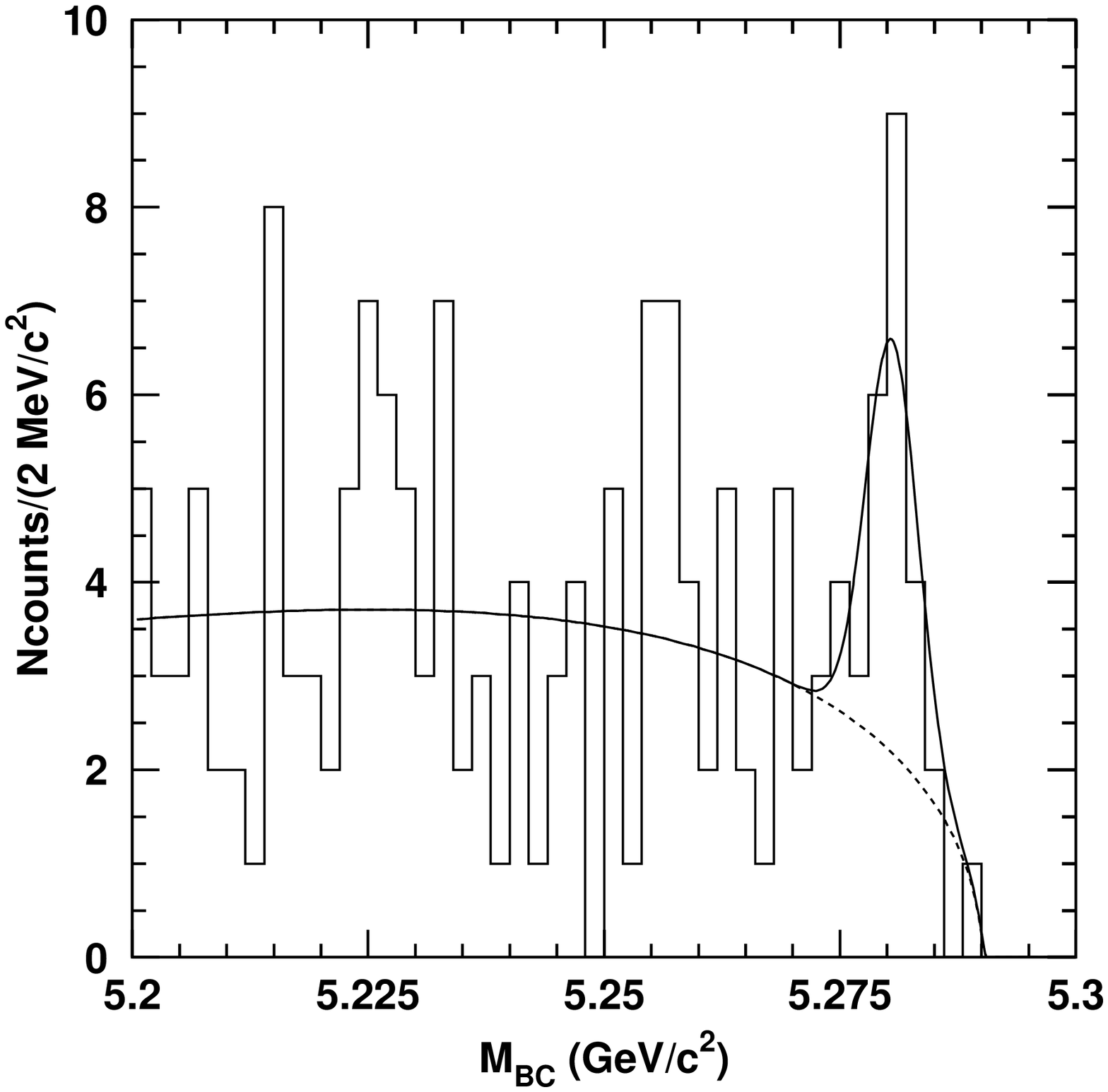} \\
    \includegraphics[height=4.0cm,width=8.0cm]{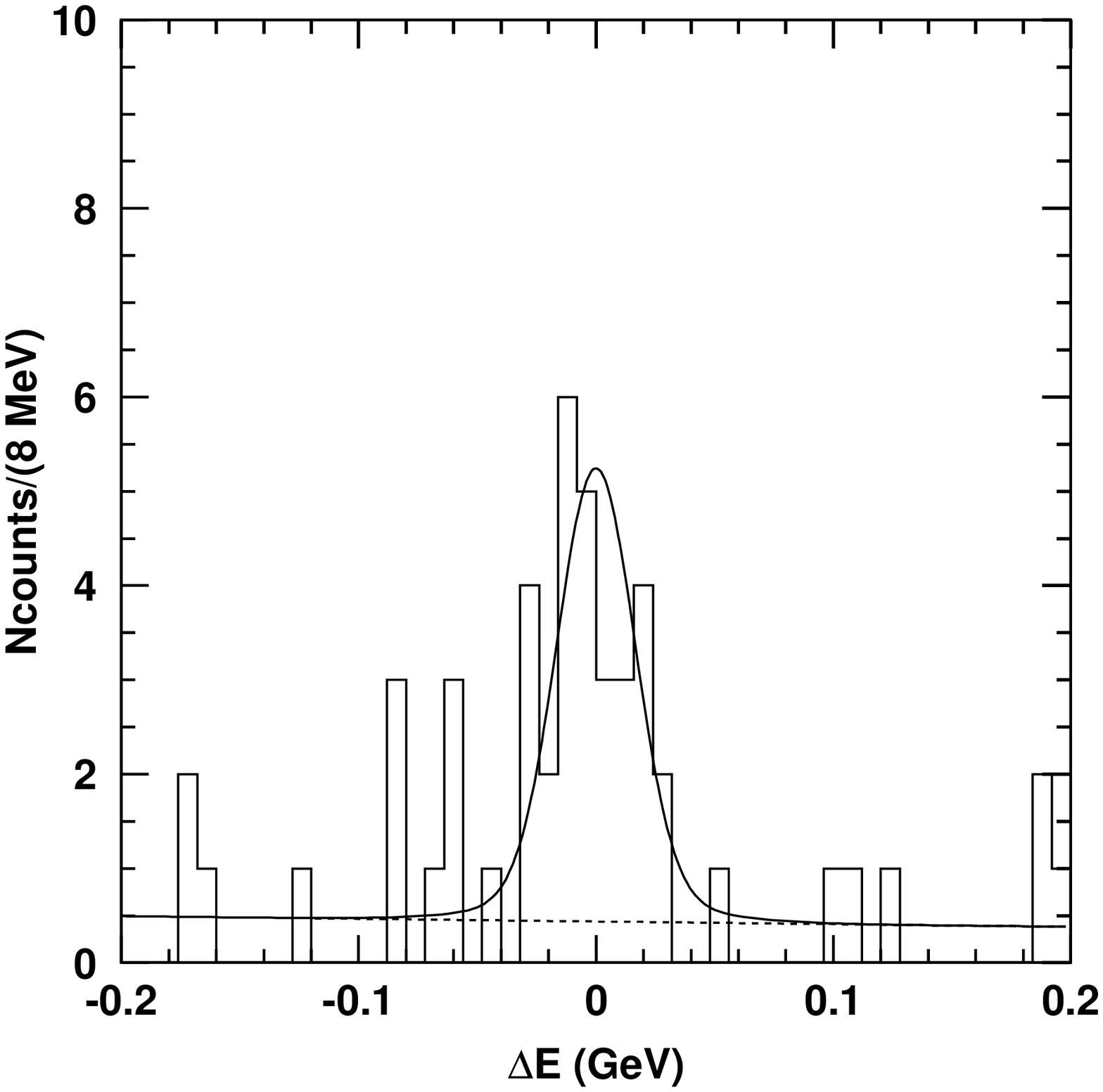} \hfill
    \includegraphics[height=4.0cm,width=8.0cm]{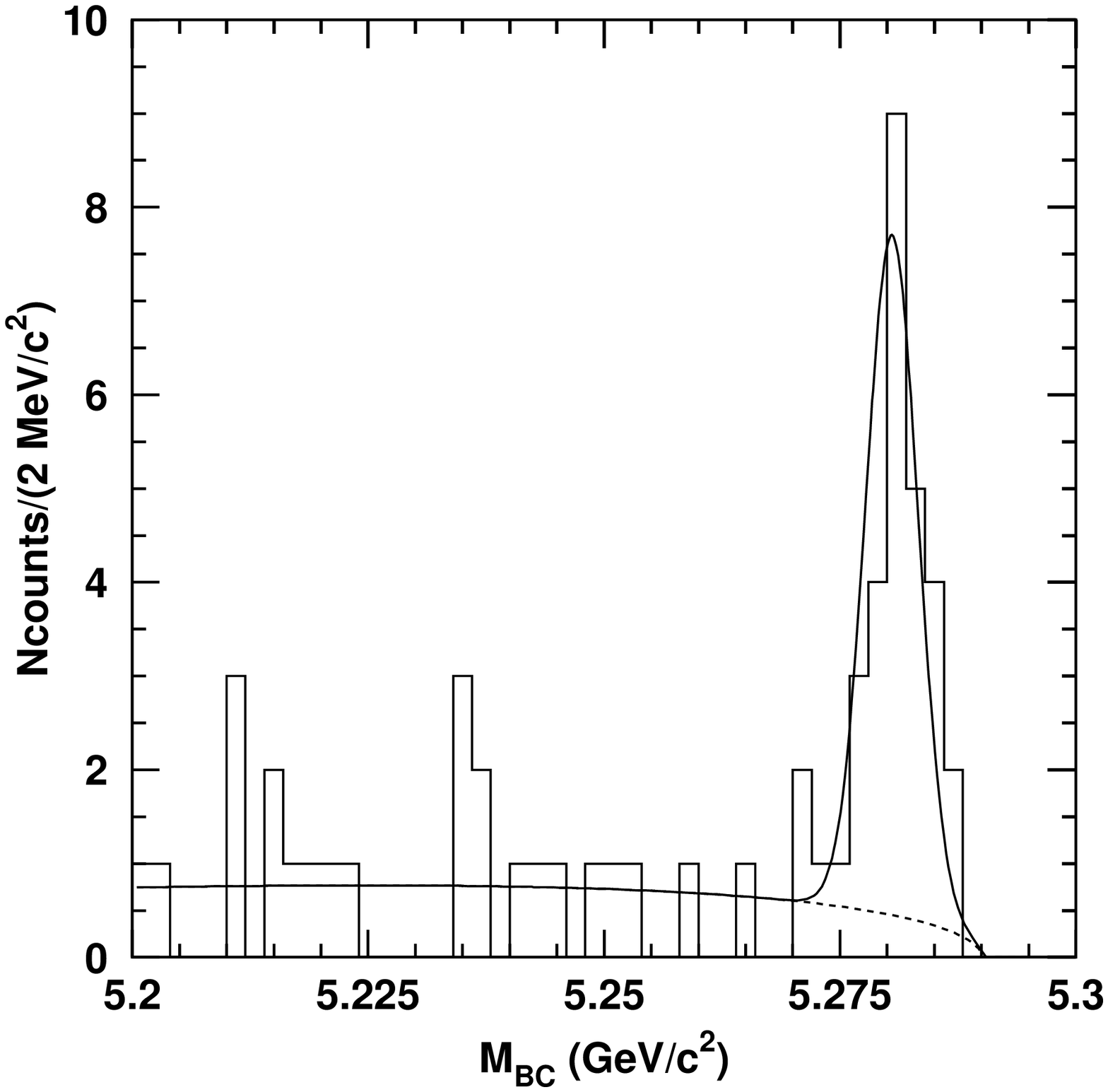}

  \caption{The $\Delta E$ (left) and $M_{BC}$ (right) distributions
           for different regions of the $K^+K^+K^-$ Dalitz plot.
           The plots from the top to bottom correspond to regions I to IV, respectively.}
  \label{kk_reso}
\end{figure}

%%%%%%%%%%%%%%%%%%%%%%%%%%%%%%%%%%%%%%%%%%%%%%%%%%%%%%%%%%%%%%%%%%%%%%%%%%%%%%%

\begin{thebibliography}{99}

\bibitem{cleokpi}{G.~Godang {\it et al.} (CLEO Collaboration),
        Phys. Rev. Lett. {\bf 80}, 3456 (1998).\\
        C.P.~Jessop {\it et al.} (CLEO Collaboration),
        Phys. Rev. Lett. {\bf 85}, 2881 (2000).\\
        D.~Cronin-Hennessy {\it et al.} (CLEO Collaboration),
        Phys. Rev. Lett. {\bf 85}, 515 (2000).}
%
\bibitem{NIM}{K.~Abe {\it et al.} (Belle Collaboration),
	KEK Progress Report 2000-4 (2000),
	to be published in Nucl. Inst. and Meth. A.}
%
\bibitem{KEKB}{KEKB B Factory Design Report, KEK Report 95-7 (1995),
	unpublished; Y. Funakoshi {\it et al.}, Proc. 2000
        European Particle Accelerator Conference, Vienna (2000).}
%
\bibitem{CDC}{H.~Hirano {\it et al.}, KEK Preprint 2000-2,
submitted to Nucl. Inst. Meth.; M. Akatsu {\it et al.},
DPNU-00-06, submitted to Nucl. Inst. Meth.}
%
\bibitem{TOF}{H.~Kichimi {\it et al.}, Nucl. Inst. Meth. A {\bf 453}, 315 (2000).}
%
\bibitem{ACC}{T.~Iijima {\it et al.}, Proceedings
of the 7th International Conference on Instrumentation for
Colliding Beam Physics, Hamamatsu, Japan, Nov 15-19, 1999.}
%
\bibitem{CSI}{H.~Ikeda {\it et al.}, Nucl. Inst. Meth. A {\bf 441}, 401 (2000).}
%
\bibitem{ARGUS}{H. Albrecht {\it et al.} (ARGUS Collaboration),
        Phys. Lett. B {\bf 229}, 304 (1989).}
%
\bibitem{VCal}{D.M.~Asner {\it et al.} (CLEO Collaboration),
        Phys. Rev. D {\bf 53}, 1039 (1996).}
%
\bibitem{Fisher}{R.A.~Fisher, Ann. Eugenics {\bf 7}, 179 (1936); \\
         M.G.~Kendall and A.~Stuart,
        {\it The Advanced Theory of Statistics}, 2nd ed. 
        (Hafner Publishing, New York, 1968), Vol. III. }
%
\bibitem{chi_c0}{K.~Abe~{\it et al.} (Belle Collaboration), BELLE-CONF-0138;
         Submitted as a contribution paper to LP2001.}
%
\bibitem{dcpk}{K.~Abe~{\it et al.} (Belle Collaboration), BELLE-CONF-0108;
         Submitted as a contribution paper to LP2001.}
%
%\bibitem{b2dk}{K.~Abe~{\it et al.} (Belle Collaboration), 
%         hep-ex/0104051, submitted to Phys. Rev. Lett.}
%
\bibitem{PDG}{D.E. Groom {\it et al.} (Particle Data Group),
	 Eur. Phys. J. C {\bf 15}, 1 (2000).
} 
%
\bibitem{b2kstp}{K.~Abe~{\it et al.} (Belle Collaboration), BELLE-CONF-0115;
         Submitted as a contribution paper to LP2001.}
%
\bibitem{b2phik}{K.~Abe~{\it et al.} (Belle Collaboration), BELLE-CONF-0113;
         Submitted as a contribution paper to EPS(2001) and LP2001.}
%
\bibitem{berg96b}{T.~Bergfeld {\it et al.} (CLEO Collaboration),
         Phys. Rev. Lett. {\bf 77}, 4503 (1996).}
%
\bibitem{babar}{T.J.~Champion (BaBar Collaboration), Proc. of the XXXth 
         Int. Conf. on High Energy Phys., Osaka (2000).}
%
\bibitem{OPAL}{G.~Abbiendi {\it et al.} (OPAL Collaboration),
         Phys. Lett. B {\bf 476}, 233 (2000).}
%
%\bibitem{buras}{G.~Buchalla, A.J.~Buras and M.E.~Lautenbacher, 
%        {\em Rev. Mod. Phys.} {\bf 68}, 1125 (1996).}
%
\bibitem{chernyak}{V.L.~Chernyak, hep-ph/0102217, 2001.}
%
\bibitem{maltman}{M.~Diehl and G.~Hiller, hep-ph/0105194.}
%\bibitem{maltman}{K.~Maltman, Phys. Lett. B {\bf 462}, 14 (1999).}
%
%\bibitem{narison}{S.~Narison, {\it QCD Spectral Sum Rules}, Lecture notes in
%         physics, Vol.~26, World Scientific, Singapore, 1989.}
%
\bibitem{anisov}{V.V.~Anisovich {\it et al.}, hep-ph/0102338, 2001.}
%
\bibitem{kim}{C.S.~Kim {\it et al.}, hep-ph/0101292, 2001.}


\end{thebibliography}
\end{document}